\documentclass[11pt,twoside]{article}
\usepackage[pdftex]{graphicx}
\usepackage{amsmath}
\usepackage{mathrsfs}
\usepackage{amssymb}
\usepackage{cite}
\usepackage{color,hyperref}
\hypersetup{urlcolor=blue}

\begin{document}

\newcommand{\pst}{\hspace*{1.5em}}

\newcommand{\rigmark}{\em Journal of Russian Laser Research}
\newcommand{\lemark}{\em Volume 30, Number 5, 2009}

%\lhead[\fancyplain{\rigmark, {\em \lemark}}{\rigmark}]{\fancyplain{\rigmark, {\em \lemark}}{\lemark}}
%\chead{}\rhead[\fancyplain{}{\lemark}]{\fancyplain{}{\rigmark}}
%\plainfootrulewidth 0.4pt
\newcommand{\be}{\begin{equation}}
\newcommand{\ee}{\end{equation}}
\newcommand{\bm}{\boldmath}
\newcommand{\ds}{\displaystyle}
\newcommand{\bea}{\begin{eqnarray}}
\newcommand{\eea}{\end{eqnarray}}
\newcommand{\ba}{\begin{array}}
\newcommand{\ea}{\end{array}}
\newcommand{\arcsinh}{\mathop{\rm arcsinh}\nolimits}
\newcommand{\arctanh}{\mathop{\rm arctanh}\nolimits}
\newcommand{\bc}{\begin{center}}
\newcommand{\ec}{\end{center}}

\thispagestyle{plain}

\label{sh}

%\lfoot[\fancyplain{\ \\[1mm] \thepage}{\ \\[1mm]\thepage}]{\fancyplain{}{}}

\begin{center} {\Large \bf
\begin{tabular}{c}
INVERSE SPIN}-{\LARGE{$\boldsymbol{{\it s}}$}}
{\Large \bf PORTRAIT\\[-1mm] AND REPRESENTATION
OF QUDIT STATES\\[-1mm] BY SINGLE PROBABILITY
VECTORS\\[-1mm]
\end{tabular}
 } \end{center}

\bigskip

\begin{center} {\bf
Sergey N. Filippov$^{1}$ and Vladimir I. Man'ko$^{2}$ }
\end{center}

\medskip

\begin{center}
{\it
$^{1}$Moscow Institute of Physics and Technology (State University)\\
Institutskii per. 9, Dolgoprudnyi, Moscow Region 141700, Russia\\
\smallskip
$^2$P.~N.~Lebedev Physical Institute, Russian Academy of Sciences\\
Leninskii Prospect 53, Moscow 119991, Russia}

\smallskip

E-mail: \href{mailto:filippovsn@gmail.com}{filippovsn@gmail.com},
\href{mailto:manko@sci.lebedev.ru}{manko@sci.lebedev.ru}
%\end{center}
%*Corresponding author, e-mail:~~~filippovsn~@~gmail.com\\
%e-mail:~~~manko~@~sci.lebedev.ru
\end{center}

\begin{abstract}\noindent
\quad Using the tomographic probability representation of qudit
states and the inverse spin-portrait method, we suggest a
bijective map of the qudit density operator onto a single
probability distribution. Within the framework of the approach
proposed, any quantum spin-$j$ state is associated with the
$(2j+1)(4j+1)$-dimensional probability vector whose components are
labeled by spin projections and points on the sphere $S^2$. Such a
vector has a clear physical meaning and can be relatively easily
measured. Quantum states form a convex subset of the $2j(4j+3)$
simplex, with the boundary being illustrated for qubits ($j=1/2$)
and qutrits ($j=1$). A relation to the $(2j+1)^2$- and
$(2j+1)(2j+2)$-dimensional probability vectors is established in
terms of spin-$s$ portraits. We also address an auxiliary problem
of the optimum reconstruction of qudit states, where the
optimality implies a minimum relative error of the density matrix
due to the errors in measured probabilities.
\end{abstract}


\section{\label{introduction}Introduction}

~~~ In the early years of quantum mechanics, it was proposed by
Landau \cite{landau} and von Neumann \cite{von-neumann} to
represent the quantum states by the density matrices. This
approach turned out to be applicable to spin states as
well~\cite{fano}. The density matrix formalism proved to be very
useful and all physical laws such as the time evolution and energy
spectrum were formulated in terms of this notion. However, many
alternative ways to describe a spin-$j$ state were proposed; for
instance, with the help of different discrete Wigner functions
(these and other analytical representations were reviewed in
\cite{vourdas}) and a fair probability-distribution function
$w(m,{\bf n})$ called spin
tomogram~\cite{dodonovPLA,oman'ko-jetp}. The latter function
depends on the spin projection $m$ along all possible unit vectors
${\bf n}\in S^2$. All these proposals are, in fact, merely
different mappings of the density operator. Inverse mappings are
also developed thoroughly and are based on the fact that, if a
(quasi)probability distribution is given, the density matrix can
be uniquely determined. On the other hand, there is a redundancy
of information contained in spin tomogram $w(m,{\bf n})$. An
attempt to avoid such a redundancy was made in
\cite{newton,amiet-weigert-JPA,amiet-weigert-JOB-L,amiet-weigert-JOB,
amiet-weigert-JPA-pure,amiet-weigert-JPA-wrong,weigert-PRA-92,weigert-PRL,heiss-weigert-PRA}.
According to \cite{newton}, the density matrix can be determined
by measuring probabilities to obtain spin projection
$m=-j,-j+1,\dots,j$ if a Stern-Gerlach apparatus is oriented along
$4j+1$ specifically chosen directions in space. In
\cite{amiet-weigert-JPA} it was shown that the density matrix can
also be reconstructed if the probabilities to get the highest spin
projection $m=j$ are known for $(2j+1)^2$ appropriately chosen
directions in space. In other words, one deals with the values of
function $w(j,{\bf n}_k)$, $k=1,2,\dots,(2j+1)^2$ and solves a
system of linear equations to express the density matrix elements
in terms of the probabilities $w(j,{\bf n}_k)$. The conditions on
vectors $\{{\bf n}_k\}_{k=1}^{(2j+1)^2}$ and an inverse method
were also presented~\cite{amiet-weigert-JPA}.

In this paper, we address the problem of identification of a
qudit-$j$ state with a single probability distribution vector.
Moreover, such a vector must have a clear physical interpretation.
These arguments make this problem interesting from both
theoretical and practical points of view. An attempt to construct
a bijective map of the density operator onto a probability vector
with some interpretation of vector components was made in the
series of
papers~\cite{fuchs-oct-2009,fuchs-07-phys-SIC,fuchs-found-phys,fuchs-qic,fuchs-sasaki}
(see also the recent review~\cite{fuchs-2010}). The problem was
shown to have an explicit solution for the lower values of spin
$j$.

In \cite{man'ko-sudarshan}, a unitary spin tomography was
suggested for describing spin states by probability distribution
functions $w(m,u)$, where $u$ is a unitary $(2j+1)\times (2j+1)$
matrix. Information contained in this probability distribution
function (called unitary spin tomogram) is even more redundant
than that contained in spin tomogram $w(m,{\bf n})$. Nevertheless,
these redundancies can be used to solve explicitly the problem
under consideration, i.e., to find an invertible map of the spin
density operator onto a probability vector with a clear physical
interpretation for an arbitrary spin $j$. The aim of our work is
to present a construction of the following invertible map. It
provides the possibility to identify any spin state with the
probability vector $\boldsymbol{\cal P}$ with components
$\mathcal{P}(m,u_k)$. Here, random variables $m$ and $u_k$ are the
spin projection and unitary matrix, respectively. The spin
projection takes values $m=-j,-j+1,\dots,j$ and a finite set of
unitary matrices $\{u_k\}_{k=1}^{N_u}$ describes the unitary
rotation operations in the finite-dimensional Hilbert space of
spin states.

Thus, the function $\mathcal{P}(m, u_k)$ is the joint probability
distribution of two random variables $m$ and $u_k$. The function
$\mathcal{P}(m,u_k)$ is related to the spin unitary tomogram
$w(m,u_k)$ by the formula
\[
w(m,u_k) = \frac{ \mathcal{P}(m,u_k) }{ \sum_{m=-j}^{j}
\mathcal{P}(m,u_k) }.
\]
\noindent This fact makes it possible to determine the density
operator $\hat{\rho}$. On the other hand, the probability
distribution $\mathcal{P}(m,u_k)$ contains extra information in
comparison with that contained in the density matrix. The matter
is that, if the density matrix $\rho$ is given, a formula for the
probability distribution $\mathcal{P}(m,u_k)$ in terms of $\rho$
does not exist. To obtain such a formula, one needs to take into
account some additional assumptions. For example, we can assume
the uniformity of the probability distribution of unitary
rotations $u_k$, i.e., each matrix $u_k$ (or direction ${\bf n}_k$
in the case $u\in SU(2)$) is taken with the same probability
$1/N_u$, where $N_u$ is an appropriate number of unitary
rotations. If this is the case, the density matrix $\rho$ provides
an explicit formula of the probability distribution ${\cal P}_{\rm
eq}(m,u_k)$. One can choose another nonuniform distribution of
unitary rotations, but knowledge of this distribution is necessary
information to be added to information contained in the density
matrix.

In the context of probability theory, the vector
$\boldsymbol{\mathcal{P}}$ (or the point on a simplex determined
by this vector) is defined by the set of spin projections $m$ and
unitary rotations $u_k$, which can be chosen with some
probability. In view of this, the probability $\mathcal{P}(m,u_k)$
under consideration is a fair joint probability distribution
function of two discrete random variables.

This paper is organized as follows.

In Sec. \ref{section:unitary:spin}, a short review of spin and
unitary spin tomograms is presented and a spin-$s$ portrait method
is introduced. In Sec. \ref{section:weigert}, we review a density
matrix reconstruction procedure proposed in
\cite{amiet-weigert-JPA}. In Sec. \ref{section:fuchs}, a
representation of quantum states by probability vectors of special
form \cite{fuchs-2010} is given. In Sec.
\ref{section:inverse-qubit-portrait}, an inverse spin-portrait
method is presented. This method allows to construct a map of the
spin density operator $\hat{\rho}$ onto the probability vector
$\boldsymbol{\cal P}$. Two cases of used unitary rotations are
given: $u\in SU(2)$ and $u\in SU(N)$, $N=2j+1$. In Sec.
\ref{section:reconstruction}, the inverse map $\boldsymbol{\cal
P}\rightarrow \hat{\rho}$ is presented in explicit form for
$SU(2)$ rotations. In Sec. \ref{section:kernels}, particular
properties of symbols ${\cal P}(m,{\bf n})$ are analyzed: star
product kernel is presented and relation to symbols $w(m,{\bf n})$
is considered. In Sec. \ref{section:examples}, examples of qubits
($j=1/2$) and qutrits ($j=1$) are presented. In Sec.
\ref{section:quantum-states}, the probability vector
$\boldsymbol{\cal P}$ is considered on the corresponding simplex
and a boundary of quantum states is presented for qubits and
qutrits. In Sec. \ref{section:conclusions}, conclusions are
presented.

\section{\label{section:unitary:spin} Unitary Spin Tomography and Spin-$\textit{s}$ Portrait of Tomograms}

~~~ We begin with some notation. Unless stated otherwise, qudit
states with spin $j$ are considered. Any state vector of such a
system is uniquely determined through the basis vectors
$|jm\rangle$, which are eigenvectors of both angular momentum
operator $\hat{J}_z$ and square of total angular momentum, i.e.,
$\hat{\bf J}^2=\hat{J}_{x}^{2}+\hat{J}_{y}^{2}+\hat{J}_{z}^{2}$.
The spin projection $m$ takes the values $-j,-j+1,\dots,j$.

A unitary spin tomogram of the state given by the density operator
$\hat{\rho}$ is defined as follows:
\begin{equation}
\label{unitary-tomogram} w^{(j)}(m, u) = \langle jm |
\hat{u}^{\dag} \hat{\rho} \hat{u} | jm \rangle = {\rm Tr} \Big(
\hat{\rho} ~ \hat{u} |j m \rangle \langle jm | \hat{u}^{\dag}
\Big) = {\rm Tr} \Big( \hat{\rho} ~ \hat{U}^{(j)}(m, u) \Big),
\end{equation}
\noindent where, in general, $\hat{u}$ is a unitary transform of
the group $SU(N)$. For the sake of convenience, starting from now
we will identify $\hat{u}$ and its matrix representation $u$ in
the basis of states $|jm\rangle$, assuming that the matrix $u$
defines, in fact, the unitary transform $\hat{u}$. The tomogram
$w^{(j)}(m, u)$ is a function of the discrete variable $m$ and the
continuous variable $u$. The operator $\hat{U}^{(j)}(m, u) = u |j
m \rangle \langle jm | u^{\dag}$ is called the dequantizer
operator because it maps an arbitrary density operator
$\hat{\rho}$ onto the real probability distribution function
$w^{(j)}(m, {\bf n})$. The dequantizer satisfies a sum rule of the
form $\sum_{m=-j}^{j}\hat{U}^{(j)}(m, u)=\hat{I}$ for all $u$. In
view of this fact, the tomogram $w^{(j)}(m, u)$ is normalized,
i.e., $\sum_{m=-j}^{j}w^{(j)}(m, u)=1$.

The particular case $u \in SU(2)$ leads to the so-called spin
tomogram $w^{(j)}(m, {\bf n})$, where the direction ${\bf n}
\equiv {\bf n}(\theta,\phi) = (\cos\phi\sin\theta,
\sin\phi\sin\theta, \cos\theta)$ determines the dequantizer
operator $\hat{U}^{(j)}(m,{\bf n})=\hat{R}({\bf n}) |j m \rangle
\langle jm | \hat{R}^{\dag}({\bf n})$  (some properties of spin
tomogram $w^{(j)}(m, {\bf n})$ were discussed in
\cite{andreev:manko,safonov:manko,andreev:safonov:manko}). Here,
we introduced a rotation operator $\hat{R}({\bf n})$ defined
through
\begin{equation}
\hat{R}({\bf n}) = e^{- i ({\bf n}_{\bot} \cdot \ \hat{\bf J})
\theta }, \qquad {\bf n}_{\bot} = (-\sin\phi, \cos\phi, 0).
\end{equation}

The inverse mapping of spin tomogram $w^{(j)}(m, {\bf n})$ onto
the density operator $\hat{\rho}$ is relatively easily expressed
through the quantizer operator $\hat{D}^{(j)}(m,{\bf n})$ as
follows:
\begin{equation}
\label{rho-reconstr-tomographic} \hat{\rho} =
\sum\limits_{m=-j}^{j} \frac{1}{4\pi} \int\limits_{0}^{2\pi} d\phi
\int\limits_{0}^{\pi} \sin \theta d\theta ~ w^{(j)}(m,{\bf
n}(\theta,\phi)) \hat{D}^{(j)}(m,{\bf n}(\theta,\phi)).
\end{equation}
\noindent Different explicit formulas of both dequantizer and
quantizer operators are
known~\cite{dodonovPLA,oman'ko-jetp,oman'ko-97,castanos,d'ariano:paini},
with the ambiguity being allowed for the quantizer operator. In
this paper, we preferably use the orthogonal expansion of the
form~\cite{filipp-spin-tomography}
\begin{eqnarray}
&& \label{dequant-orth} \hat{U}^{(j)}(m,{\bf n}) = \sum_{L=0}^{2j}
f_L^{(j)}(m) \hat{R}({\bf n}) \hat{S}_{L}^{(j)}
\hat{R}^{\dag}({\bf n}) = \sum_{L=0}^{2j}
f_L^{(j)}(m) \hat{S}_{L}^{(j)}({\bf n}), \\
&& \label{quant-orth} \hat{D}^{(j)}(m,{\bf n}) = \sum_{L=0}^{2j}
(2L+1) f_L^{(j)}(m) \hat{R}({\bf n}) \hat{S}_{L}^{(j)}
\hat{R}^{\dag}({\bf n}) = \sum_{L=0}^{2j} (2L+1) f_L^{(j)}(m)
\hat{S}_{L}^{(j)}({\bf n}),
\end{eqnarray}
\noindent where the coefficient $f_L^{(j)}(m)$ is an $L$-degree
polynomial of the discrete variable $m$, and the operator
$\hat{S}_{L}^{(j)}({\bf n})$ is the same polynomial of the
operator variable $\hat{R}({\bf n}) \hat{J}_z \hat{R}^{\dag}({\bf
n})= (\hat{\bf J}\cdot {\bf n})$.

For instance, in the case of qubits ($j=1/2$) and qutrits ($j=1$),
we have
\begin{eqnarray}
&& f_0^{(1/2)}(m)= \frac{1}{\sqrt{2}}, \quad f_1^{(1/2)}(m)=
\sqrt{2} m, \quad \hat{S}_0^{(1/2)}({\bf n})=
\frac{1}{\sqrt{2}}\hat{I}, \quad \hat{S}_1^{(1/2)}({\bf n})=
\sqrt{2} (\hat{\bf J}\cdot {\bf n}),\\
&& f_0^{(1)}(m)= \frac{1}{\sqrt{3}}, \quad f_1^{(1)}(m)=
\frac{m}{\sqrt{2}}, \quad f_2^{(1)}(m)= \frac{3m^2-2}{\sqrt{6}},\\
&& \hat{S}_0^{(1)}({\bf n})= \frac{1}{\sqrt{3}}\hat{I}, \quad
\hat{S}_1^{(1)}({\bf n})= \frac{1}{\sqrt{2}}(\hat{\bf J}\cdot {\bf
n}), \quad \hat{S}_2^{(1)}({\bf n})= \frac{1}{\sqrt{6}}\left( 3
(\hat{\bf J}\cdot {\bf n})^2 - 2 \hat{I} \right).
\end{eqnarray}

In the case of an arbitrary spin $j$,
$f_0^{(j)}(m)=1/\sqrt{2j+1}$,
$f_1^{(j)}(m)=\sqrt{3}m/\sqrt{j(j+1)(2j+1)}$, and all other
coefficients $f_L^{(j)}(m)$ are expressed via the recurrence
relation that was presented in \cite{filipp-spin-tomography} and
relates $f_L^{(j)}(m)$, $f_{L-1}^{(j)}(m)$, and
$f_{L-2}^{(j)}(m)$. Expansions~(\ref{dequant-orth}) and
(\ref{quant-orth}) are orthogonal in the following sense:
\begin{equation}
{\rm Tr} \left( \hat{S}_{L}^{(j)}({\bf n}) \hat{S}_{L'}^{(j)}({\bf
n}) \right) = \sum \limits_{m=-j}^{j} f_L^{(j)}(m) f_{L'}^{(j)}(m)
= \delta_{LL'}.
\end{equation}
\noindent This means that functions $f_L^{(j)}(m)$ are orthogonal
polynomials of a discrete variable, with the weight function being
identically equal to unity. Using the theory of classical
orthogonal polynomials of a discrete
variable~\cite{nikiforov-suslov-uvarov,nikiforov-uvarov}, it is
not hard to prove that the function $f_L^{(j)}(m)$ is expressed
through the discrete Chebyshev polynomial $t_n(x,N)$ or Hahn
polynomial $h_n^{(\alpha,\beta)}(x,N)$ as follows:
\begin{equation}
f_L^{(j)}(m) = \frac{1}{d_L} t_L(j+m,2j+1) = \frac{1}{d_L}
h_L^{(0,0)}(j+m,2j+1), \qquad d_L =
\sqrt{\frac{(2j+L+1)!}{(2L+1)(2j-L)!}} .
\end{equation}

\subsection{\label{subsection:spin-s-portrait}Spin-$\textit{s}$ Portrait}

~~~ The tomogram $w^{(j)}(m,u)$ of a system with spin $j$ is a
function of the discrete spin projection $m$. This means that the
tomogram $w^{(j)}(m,u)$ can be represented in the form of the
following $(2j+1)$-dimensional probability vector:
\begin{equation}
\label{portrait-original}
{\bf w}_{j}^{(j)}(u) = \left(%
\begin{array}{c}
  w^{(j)}(j,u) \\
  w^{(j)}(j-1,u) \\
  \cdots \\
  w^{(j)}(-j,u) \\
\end{array}%
\right).
\end{equation}
\noindent We will refer to such a $(2j+1)$ vector as the spin-$j$
portrait (in analogy with the qubit-portrait
concept~\cite{chernega}). Note, that
vector~(\ref{portrait-original}) is a fair probability
distribution vector since $w(m,u) \ge 0$ and
$\sum_{m=-j}^{j}w(m,u)=1$ for all unitary matrices $u$. Since the
components of vector~(\ref{portrait-original}) are fair
probabilities, the sum $\sum_{m \in {\cal A}} w^{(j)}(m,u)$, where
${\cal A} \subset \{ m \}_{-j}^{j}$, can also be treated as a
probability and has a clear physical meaning. Summing some
components of vector~(\ref{portrait-original}), one can construct
a probability vector ${\bf w}_{s}^{(j)}$ of less dimension
$(2s+1)$, where $s$ plays the role of pseudospin and can take
values $s=1/2,1,\dots,j-1/2,j$. To be precise, ${\bf w}_{s}^{(j)}$
reads
\begin{equation}
\label{portrait-spin-s}
{\bf w}_{s}^{(j)}(u) = \left(%
\begin{array}{c}
  \sum_{m \in {\cal A}_1} w^{(j)}(m,u) \\
  \sum_{m \in {\cal A}_2} w^{(j)}(m,u) \\
  \cdots \\
  \sum_{m \in {\cal A}_{2s+1}} w^{(j)}(m,u) \\
\end{array}%
\right),
\end{equation}
\noindent where ${\cal A}_k \subset \{m\}_{-j}^{j}$ for all
$k=1,\dots,2s+1$, $\cup_{k=1}^{2s+1}{\cal A}_k = \{m\}_{-j}^{j}$,
and ${\cal A}_k \cap {\cal A}_l = \emptyset$ for all $k \ne l$.
Vector~(\ref{portrait-spin-s}) is referred to as the spin-$s$
portrait of a qudit-$j$ state. In the case $s=1/2$, we obtain the
so-called qubit portrait of the form
\begin{eqnarray}
\label{portrait-qubit}
& {\bf w}_{1/2}^{(j)}(u) = \left(%
\begin{array}{c}
  \sum_{m \in {\cal A}_1} w^{(j)}(m,u) \\
  \sum_{m \in {\cal A}_2} w^{(j)}(m,u) \\
\end{array}%
\right) = \left(%
\begin{array}{c}
  \sum_{m \in {\cal A}_1} w^{(j)}(m,u) \\
  1-\sum_{m \in {\cal A}_1} w^{(j)}(m,u) \\
\end{array}%
\right) = \left(%
\begin{array}{c}
  1-\sum_{m \in {\cal A}_2} w^{(j)}(m,u) \\
  \sum_{m \in {\cal A}_2} w^{(j)}(m,u) \\
\end{array}%
\right), \nonumber \\
&
\end{eqnarray}
\noindent with ${\cal A}_1 \cup {\cal A}_2= \{m\}_{-j}^{j}$,
${\cal A}_1 \cap {\cal A}_2 = \emptyset$. In the particular case
of ${\cal A}_1=\{j\}$ and ${\cal A}_2=\{m\}_{-j}^{j-1}$, the qubit
portrait~(\ref{portrait-qubit}) reads
\begin{equation}
\label{portrait-qubit-projection-j}
{\bf w}_{1/2}^{(j)}(u) = \left(%
\begin{array}{c}
  w^{(j)}(j,u) \\
  \sum_{m = -j}^{j-1} w^{(j)}(m,u) \\
\end{array}%
\right) = \left(%
\begin{array}{c}
  w^{(j)}(j,u) \\
  1- w^{(j)}(j,u) \\
\end{array}%
\right).
\end{equation}
\noindent Qubit portraits of this kind are implicitly used in
several reconstruction procedures considered in subsequent
sections. The qubit-portrait method is introduced in
\cite{chernega} and successfully applied not only to spin
systems~\cite{lupo} but also to the light
states~\cite{filipp-qubit-portrait}.

The inverse spin-portrait method is to construct a single
probability distribution vector with the help of several spin-$s$
portraits. Indeed, one can stack $N_s$ spin-$s$ portraits ${\bf
w}_{s}^{(j)}(u)$ into the following single probability vector of
final dimension $N_s(2s+1)$:
\begin{equation}
\label{inverse-spin-s-portrait}
\frac{1}{N_s} \left(%
\begin{array}{c}
  {\bf w}_{s}^{(j)}(u_1) \\
  {\bf w}_{s}^{(j)}(u_2) \\
  \cdots \\
  {\bf w}_{s}^{(j)}(u_{N_s}) \\
\end{array}%
\right).
\end{equation}
\noindent This idea is elaborated for the case $s=j$ in Sec.
\ref{section:inverse-qubit-portrait}. Here, we restrict ourselves
to the discussion of a particular case $s=1/2$ (qubit portrait).
Only one component of two-vector~(\ref{portrait-qubit}) contains
information on the system. The density operator $\hat{\rho}$ is
determined by $(2j+1)^2$ real numbers without regard to the
normalization condition. This fact shows that, if a single
probability distribution~(\ref{inverse-spin-s-portrait}) contains
complete information on the system, then it should comprise at
least $N_{1/2}=(2j+1)^2$ different qubit portraits. Then the
dimension of vector~(\ref{inverse-spin-s-portrait}) is
$2(2j+1)^2$, but only $(2j+1)^2$ components are relevant. In the
following section, we review the reconstruction procedure that was
suggested in \cite{amiet-weigert-JPA} and that employed $(2j+1)^2$
qubit portraits of the form (\ref{portrait-qubit-projection-j})
with $u_k \in SU(2)$, $k=1,\dots,(2j+1)^2$. The symmetric
informationally complete POVM (positive operator-valued measure)
to be outlined in Sec.~\ref{section:fuchs} can also be treated as
an implicit use of $(2j+1)^2$ qubit
portraits~(\ref{portrait-qubit-projection-j}) with unitary
matrices $u_k\in SU(N)$, $N=2j+1$, $k=1,\dots,(2j+1)^2$ which
satisfy additional requirements
$\sum_{k=1}^{(2j+1)^2}\hat{u}_k|jj\rangle\langle
jj|\hat{u}_k^{\dag}=(2j+1)\hat{I}$ and $\langle
jj|\hat{u}_k^{\dag}\hat{u}_l|jj\rangle=1/(2j+2)$ for all $k\ne l$.

\section{\label{section:weigert} Amiet--Weigert Reconstruction of the Density Matrix}

~~~ In this section, we review the approach
\cite{amiet-weigert-JPA} to reconstruct the density matrix of a
spin $j$ through the Stern--Gerlach measurements, where one
measures the probabilities to obtain the maximum spin projection
$m=j$ for $(2j+1)^2$ appropriately chosen directions ${\bf n}_k$
in space.

According to (\ref{unitary-tomogram}), for each fixed direction
${\bf n}_k$, $k=1,\dots,(2j+1)^2$, the probability to obtain the
spin projection $m=j$ is given by the formula
\begin{equation}
w^{(j)}(j,{\bf n}_k) = \langle jj| \hat{R}^{\dag}({\bf n}_k)
\hat{\rho} \hat{R}({\bf n}_k) |jj \rangle.
\end{equation}
\noindent We denote by {\bf W} a vector comprising all these
probabilities
\begin{equation}
\label{w-a-vector}
{\bf W} = \left(%
\begin{array}{cccc}
  w^{(j)}(j,{\bf n}_1) & w^{(j)}(j,{\bf n}_2) & \dots & w^{(j)}(j,{\bf n}_{(2j+1)^2}) \\
\end{array}%
\right)^{T}.
\end{equation}
\noindent Moreover, the $(2j+1)$$\times$$(2j+1)$ density matrix
$\rho$ can be written in the form of a $(2j+1)^2$-component vector
\begin{eqnarray}
{\boldsymbol \rho} = \left(%
\begin{array}{ccccccccccccc}
  \rho_{11} & \rho_{21} & \ldots & \rho_{2j+1,1} & \rho_{12} & \rho_{22} & \ldots & \rho_{2j+1,2} &
  \dots & \rho_{1,2j+1} & \rho_{2,2j+1} & \ldots & \rho_{2j+1,2j+1} \\
\end{array}%
\right)^{T},\nonumber\\
\end{eqnarray} \noindent which implies that
the density-matrix columns are merely stacked up in a single
column. Let ${\bf U}^{(j)}(m,{\bf n})$ be a vector constructed
from the operator $\hat{U}^{(j)}(m,{\bf n})$ using the same rule.
Then the probability $w^{(j)}(m,{\bf n})$ is nothing else but the
scalar product of two $(2j+1)^2$-component vectors
\begin{equation}
w^{(j)}(m,{\bf n}) = \left( {\bf U}^{(j)}(m,{\bf n}) \cdot
\boldsymbol\rho \right) = {\rm Tr} \left[ {\bf U}^{(j)}(m,{\bf
n})^T ~ \boldsymbol\rho \right].
\end{equation}
\noindent From this, it is not hard to see that the map
$\boldsymbol\rho \rightarrow {\bf W}$ reads
\begin{equation}
\label{w-a-W} {\bf W} = \|M\| \boldsymbol\rho,
\end{equation}
\noindent where $\|M\|$ is a $(2j+1)^2\times (2j+1)^2$ matrix of
the form
\begin{equation}
\|M\| = \left(%
\begin{array}{c}
  {\bf U}^{(j)}(j,{\bf n}_1)^{T} \\
  {\bf U}^{(j)}(j,{\bf n}_2)^{T} \\
  \cdots \\
  {\bf U}^{(j)}(j,{\bf n}_{(2j+1)^2})^{T} \\
\end{array}%
\right).
\end{equation}

Whenever $\det\|M\| \ne 0$, there exists an inverse map ${\bf W}
\rightarrow \boldsymbol\rho$ and
\begin{equation}
\label{w-a-rho} \boldsymbol\rho = \|M^{-1}\| {\bf W}.
\end{equation}
In \cite{amiet-weigert-JPA}, it is shown that a particular choice
of directions ${\bf n}_k$, $k=1,\dots,(2j+1)^2$ ensures that the
condition $\det\|M\| \ne 0$ is fulfilled. Moreover, such a choice
of directions simplifies significantly the inversion procedure
because it relies on the Fourier transform. Namely, the following
directions were proposed:
\begin{equation}
\label{W-A-directions} {\bf n}_k \equiv {\bf n}_{qr} \equiv
(\sin\theta_q \cos\varphi_{qr}, \sin\theta_q \sin\varphi_{qr},
\cos\theta_q), \qquad 0 \le q,r \le 2j,
\end{equation}
\noindent with $0 < \theta_q < \pi$, $\theta_q \ne \theta_{q'}$ if
$q \ne q',$ and
\begin{equation}
 \varphi_{qr} =
\frac{2\pi}{2j+1}(r+q\Delta), \qquad 0 < \Delta \le
\frac{1}{2j+1}.
\end{equation}

\begin{figure}[t]
\begin{center}
\includegraphics{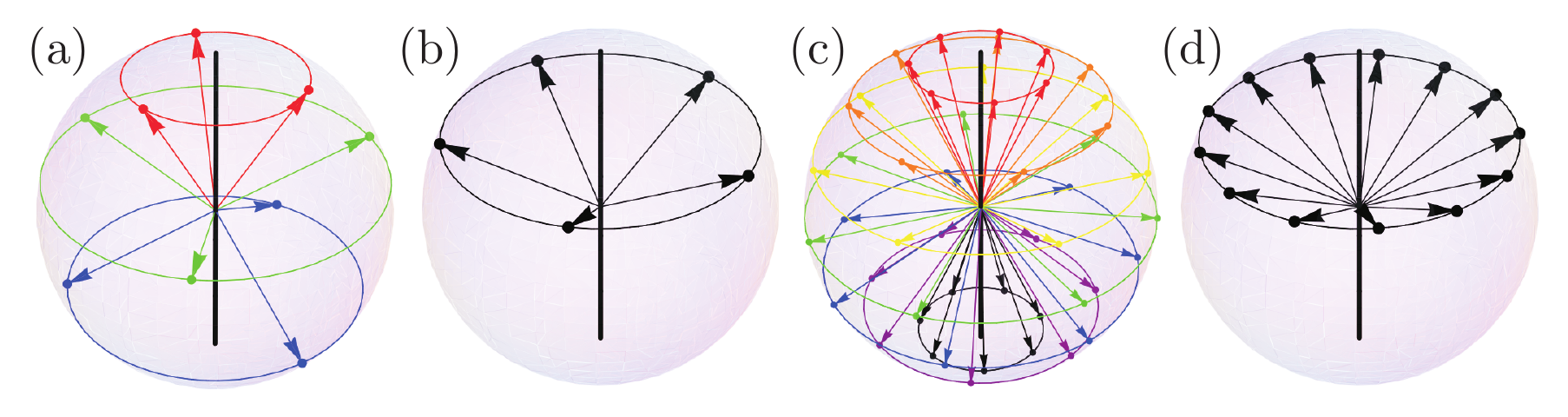}
\caption{\label{Weigert-Newton} Reconstruction of spin-$j$ density
matrices: $j=1$ (a,~b) and $j=3$ (c,~d). The procedure developed
in \cite{amiet-weigert-JPA} is represented by a and c; it is based
on fixing $(2j+1)^2$ specifically arranged
directions~(\ref{W-A-directions}) and measuring the probabilities
of the spin projection onto each direction to take the value $j$.
The method used in \cite{newton} is illustrated by b and d; in
this case, one needs to measure $4j+1$ spin-$j$ portraits ${\bf
w}_{j}^{(j)}({\bf n}_k)$, where ${\bf n}_k$ are specifically
chosen to form a cone.}
\end{center}
\end{figure}

Examples of such a choice of the directions for spins $j=1$ and
$j=3$ are shown in Fig.~\ref{Weigert-Newton}. We see that the
directions are divided into groups that form nested cones. Some
modifications to free cones and spirals were presented in
\cite{heiss-weigert-PRA}. An arbitrary choice of the directions
was discussed in \cite{amiet-weigert-JOB}.

%%%%%%%%%%%%%%%%%%%%%%%%%%%%%%%%%%%%%%%%%%%%%%%%%%%%%%%%%%%%%%%%%%%%%%%%%%%%%%%%%
%%%%%%%%%%%%%%%%%%%%%%%%%%%%%%%%%%%%%%%%%%%%%%%%%%%%%%%%%%%%%%%%%%%%%%%%%%%%%%%%%
%%%%%%%%%%%%%%%%%%%%%%%%%%%%%%%%%%%%%%%%%%%%%%%%%%%%%%%%%%%%%%%%%%%%%%%%%%%%%%%%%
%%%%%%%%%%%%%%%%%%%%%%%%%%%%%%%%%%%%%%%%%%%%%%%%%%%%%%%%%%%%%%%%%%%%%%%%%%%%%%%%%

\section{\label{section:fuchs} Quantum States as Probability Distributions}

~~~ To begin, we showed in Sec.~\ref{section:unitary:spin} that
any quantum state can be interpreted as a probability distribution
function. Since tomogram~(\ref{unitary-tomogram}) is a function
depending on the unitary matrix $u$ (direction ${\bf n}$ in the
case of $u \in SU(2)$), there is a redundancy of information
contained in the tomogram. It is tempting to reduce such a
redundancy and associate a quantum state with the single
probability distribution. The proposal to associate quantum states
with single probability vectors was made in
\cite{fuchs-2010,fuchs-oct-2009,fuchs-07-phys-SIC,fuchs-found-phys,fuchs-qic,fuchs-sasaki}.
To avoid any redundancy, it was suggested to use the minimum
informationally complete POVM (positive operator-valued measure).
This suggestion is related to constructing the minimum tomographic
set discussed in \cite{Ventri}. Using the language of spin states,
any spin-$j$ state is associated with the following
$(2j+1)^2$-component probability vector:
\begin{equation}
{\bf p} = \left(%
\begin{array}{c}
  {\rm tr}(\hat{\rho} \hat{E}_{1}) \\
  {\rm tr}(\hat{\rho} \hat{E}_{2}) \\
  \cdots \\
  {\rm tr}(\hat{\rho} \hat{E}_{(2j+1)^2}) \\
\end{array}%
\right),
\end{equation}
\noindent where $\hat{E}_{i}$, $i=1,\dots,(2j+1)^2$ are the
corresponding POVM effects. There are many ways to introduce the
minimum informationally complete POVM effects, and a possible
choice that is valid in any dimension was presented in
\cite{fuchs-found-phys}. A relatively new tendency is to use the
symmetric informationally complete (SIC) POVMs. If this is the
case, the effects $\{\hat{E}_i\}_{i=1}^{(2j+1)^2}$ are
one-dimensional projectors satisfying the following condition:
\begin{equation}
{\rm tr}(\hat{E}_k \hat{E}_l) = \frac{1}{(2j+1)^2(2j+2)} \quad
{\rm if } \quad k \ne l.
\end{equation}
\noindent As stated in \cite{fuchs-oct-2009}, the effects
$\{\hat{E}_i\}_{i=1}^{(2j+1)^2}$ that meet the above requirements
were found numerically in dimensions $(2j+1)\le 67$ and
analytically in dimensions $(2j+1)=2-15,~19,~{\rm and}~24$.

It is worth noting that the Amiet--Weigert construction considered
in the previous section can be treated as a single probability
distribution. For this, one needs to normalize
vector~(\ref{w-a-vector}). In this case, the
mappings~(\ref{w-a-W}) and (\ref{w-a-rho}) are slightly modified
\begin{equation}
{\bf W'} = \frac{\|M\|
\boldsymbol\rho}{\sum_{i,k=1}^{(2j+1)^2}\|M\|_{ik}\rho_k}, \qquad
\boldsymbol\rho = \frac{\|M\|^{-1} {\bf
W'}}{\sum_{i=1}^{2j+1}\sum_{k=1}^{(2j+1)^2}\|M\|_{(i-1)(2j+1)+i,k}^{-1}
{W'_{k}}}.
\end{equation}

\section{\label{section:inverse-qubit-portrait} Inverse Spin-Portrait Method}
In this section, starting from spin tomograms, we associate each
quantum state with the corresponding probability distribution
vector, which has a clear physical meaning and can be measured
experimentally.

The unitary spin tomogram $w^{(j)}(m,u)$ is a function of the
unitary matrix $u$ and depends on a discrete parameter
$m=-j,\dots,j-1,j$. Fixing the unitary rotation $u_k$, we obtain a
$(2j+1)$-component probability vector ${\bf w}_j^{(j)}(u_k)$,
called the spin-$j$ portrait of the system (see Sec.~2.1). Given
only one spin-$j$ portrait of the system, it is impossible, in
general, to define without doubt a state of the system.
Nevertheless, the state is determined if one has an adequate
number $N_u$ of different spin-$j$ portraits. Then we can
introduce a joint probability distribution function of two random
variables $m$ and $u_k$
\begin{equation}
\label{P-m-uk} \mathcal{P}(m,u_k), \qquad m=-j,-j+1,\dots,j, \quad
k=1,2,\dots,N_u.
\end{equation}
\noindent The physical meaning of this joint probability
distribution is that, if one randomly chooses a unitary rotation
from the set $\{u_k\}_{k=1}^{N_u}$ and a spin projection $m$
within the interval $-j \le m \le j$, the value of
$\mathcal{P}(m,u_k)$ gives the probability of the detector's
click. Function~(\ref{P-m-uk}) can also be written in the form of
the following $N_u(2j+1)$-component probability distribution
vector:
\begin{eqnarray}
\boldsymbol{\cal P} = \left(%
\begin{array}{cccccccccc}
  \mathcal{P}(j,u_1) & \ldots & \mathcal{P}(-j,u_1) & \mathcal{P}(j,u_2) & \ldots & \mathcal{P}(-j,u_2)
  & \ldots & \mathcal{P}(j,u_{N_u}) & \ldots & \mathcal{P}(-j,u_{N_u}) \\
\end{array}%
\right)^{T}\nonumber\\
\end{eqnarray}
\noindent with the normalization condition of the form
\begin{equation}
\sum_{m=-j}^{j} \sum_{k=1}^{N_u} \mathcal{P}(m,u_k) = 1.
\end{equation}

Since the tomogram $w^{(j)}(m,u_k)$ is nothing else but the
probability to obtain the spin projection $m$ if the rotation
$u_k$ is fixed, the relation between the spin tomogram and the
joint probability distribution function $\mathcal{P}(m,u_k)$ reads
\begin{equation}
\label{tom-from-P} w^{(j)}(m,u_k) = \frac{ \mathcal{P}(m,u_k) }{
\sum_{m=-j}^{j} \mathcal{P}(m,u_k) },
\end{equation}
\noindent where the denominator has the sense of the probability
$p_k$ to choose the unitary rotation $u_k$. If the probabilities
$\{p_k\}_{k=1}^{N_u}$ are known a priori, then the vector
$\boldsymbol{\cal P}$ is easily expressed via spin-$j$
portraits~(\ref{portrait-original}), namely,
\begin{equation}
\label{vector-apriori}
\boldsymbol{\cal P} = \left(%
\begin{array}{c}
  p_1 {\bf w}_{j}^{(j)}(u_1) \\
  p_2 {\bf w}_{j}^{(j)}(u_2) \\
  \cdots \\
  p_{N_u} {\bf w}_{j}^{(j)}(u_{N_u}) \\
\end{array}%
\right).
\end{equation}
\noindent If the unitary rotations $u_k$, $k=1,\dots,N_u$ are
equiprobable, then
\begin{equation}
\boldsymbol{\cal P}_{\rm eq} = \frac{1}{N_u} \left(%
\begin{array}{c}
  {\bf w}_{j}^{(j)}(u_1) \\
  {\bf w}_{j}^{(j)}(u_2) \\
  \cdots \\
  {\bf w}_{j}^{(j)}(u_{N_u}) \\
\end{array}%
\right).
\end{equation}

It is worth mentioning that, even if a priori the probabilities
$\{p_k\}_{k=1}^{N_u}$ are not known, formula~(\ref{tom-from-P})
provides a direct way of mapping $\boldsymbol{\cal P}$ onto the
vector $\boldsymbol{\cal P}_{\rm eq}$.

Let us now consider an open problem of the minimum number $N_u$ of
spin portraits. In other words, $N_u$ is the number of unitary
rotations $u_k$ that is needed to identify any quantum state with
a single probability vector of the form~(\ref{vector-apriori}) and
to minimize the redundancy of information contained in this
vector. Subsequently, two main cases are presented, namely, the
use of $SU(2)$ rotations and $SU(N)$ rotations with $N=2j+1$.
These particular problems can be of interest for experimentalists
because $SU(2)$ rotations can be relatively easily realized in
some modifications of the Stern--Gerlach experiment, while $SU(N)$
rotations may require more difficult apparatus. On the other hand,
it will be shown that, to extract information on the system, one
can use a smaller number of $SU(N)$ rotations than in the case of
$SU(2)$ matrices.

\subsection{\label{subsection:number-of-rotations} $\textit{SU(2)}$ Rotations}
~~~ Like the Amiet--Weigert scanning procedure (\ref{w-a-W}), the
map of the density operator $\hat{\rho}$ onto the probability
vector (\ref{vector-apriori}) can be written as follows:
\begin{equation}
\label{P-map} {\boldsymbol {\cal P}} = \|Q\| \boldsymbol\rho,
\end{equation}
\noindent where $\|Q\|$ is an $N_u(2j+1)\times (2j+1)^2$
rectangular matrix of the form
\begin{equation}
\|Q\| = \left(%
\begin{array}{c}
 p_1 {\bf U}^{(j)}(j,{\bf n}_1)^{T} \\
 p_1 {\bf U}^{(j)}(j-1,{\bf n}_1)^{T} \\
  \cdots \\
 p_1 {\bf U}^{(j)}(-j,{\bf n}_1)^{T} \\
 \vdots\\
 p_{N_u} {\bf U}^{(j)}(j,{\bf n}_{N_u})^{T} \\
 p_{N_u} {\bf U}^{(j)}(j-1,{\bf n}_{N_u})^{T} \\
  \cdots \\
 p_{N_u} {\bf U}^{(j)}(-j,{\bf n}_{N_u})^{T} \\
\end{array}%
\right).
\end{equation}
\noindent The map (\ref{P-map}) is invertible iff ${\rm
rank}\|Q\|=(2j+1)^2$. Since the rank of a matrix is equal to the
number of linearly independent rows, the number $N_u$ can be
defined as the minimum natural number such that the set $\{{\bf
U}^{(j)}(m,{\bf n}_k)\}$, $m=-j,\dots,j$, $k=1,\dots,N_u$ contains
$(2j+1)^2$ linearly independent vectors. According to
\cite{filipp-spin-tomography}, each vector ${\bf U}^{(j)}(m,{\bf
n})$ can be resolved to the sum of orthogonal vectors ${\bf
S}_{L}^{(j)}({\bf n})$, $L=0,\dots,2j$; namely,
\begin{equation}
{\bf U}^{(j)}(m,{\bf n}) = \sum_{L=0}^{2j}f_L^{(j)}(m){\bf
S}_L^{(j)}({\bf n}),
\end{equation}
\noindent where $\left( {\bf S}_L^{(j)}({\bf n}) \cdot {\bf
S}_{L'}^{(j)}({\bf n}) \right) = \sum_{m=-j}^{j} f_{L}^{(j)}(m)
f_{L'}^{(j)}(m) = \delta_{LL'}$. The vector ${\bf S}_L^{(j)}({\bf
n})$ corresponds to the operator $\hat{S}_L^{(j)}({\bf n})$ acting
on the Hilbert space of spin-$j$ states (see
Sec.~\ref{section:unitary:spin}). The operator
$\hat{S}_L^{(j)}({\bf n})$ is shown to be the same polynomial
$f_L^{(j)}(m)$ of degree $L$, with the argument being replaced:
$~m \longrightarrow (\hat{\bf J}\cdot{\bf n}) = \hat{J}_x n_x +
\hat{J}_y n_y + \hat{J}_z n_z$. Suppose $L=0$, then there exists
only one linear independent vector of the form ${\bf
S}_{L=0}^{(j)}({\bf n})$ which corresponds to the identity
operator $\hat{I}$. If $L=1$, no more than three linear
independent vectors ${\bf S}_{L=1}^{(j)}({\bf n}_k)$, $k=1,2,3$,
which correspond to the operators $(\hat{\bf J}\cdot{\bf n}_1)$,
$(\hat{\bf J}\cdot{\bf n}_2)$, and $(\hat{\bf J}\cdot{\bf n}_3)$,
respectively, can exist. Note that the three vectors ${\bf
S}_{L=1}^{(j)}({\bf n}_k)$, $k=1,2,3$ are independent iff vectors
${\bf n}_k \in \mathbb{R}^3$ are not coplanar, i.e., their triple
product $({\bf n}_1 \cdot [{\bf n}_2 \times {\bf n}_3] ) \ne 0$.
Taking into account the normalization conditions ${\bf n}_k^2=1$
and $\hat{\bf J}^2=j(j+1)\hat{I}$, in the case $L=2$, we obtain
five linear independent vectors ${\bf S}_{L=2}^{(j)}({\bf n}_k)$.
Using the matrix representation of the operator
$\hat{S}_{L}^{(j)}({\bf n})$, we see that it is composed of $2L+1$
independent $l$-diagonal operators ($l=0$ for a diagonal one,
$l=1$ for a super-diagonal one, $l=-1$ for a sub-diagonal one, and
so on for all $-L \le l \le L$). Increasing $L$ by unity, two more
diagonals are filled. We draw the conclusion that for a fixed $L$
the maximum number of linearly independent vectors ${\bf
S}_{L}^{(j)}({\bf n}_k)$ is equal to $2L+1$. Since vectors ${\bf
S}_{L}^{(j)}({\bf n}_k)$ and ${\bf S}_{L'}^{(j)}({\bf n}_k)$ with
different $L$ and $L'$ are orthogonal, the total number of linear
independent rows ${\bf U}_j^{T}(m,{\bf n})$ equals
$1+3+5+\dots+N_u=(N_u+1)^2/4$. On the other hand, it must be equal
to ${\rm rank}\|Q\|=(2j+1)^2$. From this, it is readily seen that
$N_u=4j+1$.

The directions ${\bf n}_k$, $k=1,\dots,4j+1$ cannot be chosen
arbitrarily because of the condition ${\rm rank}\|Q\|=(2j+1)^2$.
As was shown above, the directions $\{{\bf n}_k\}_{k=1}^{4j+1}$
are divided into sets of one, three, five, and so on directions.
Without loss of generality, it can be assumed that these sets are
$\{{\bf n}_1\}$, $\{{\bf n}_k\}_{k=1}^{3}$, $\{{\bf
n}_k\}_{k=1}^{5}$, $\dots$, $\{{\bf n}_k\}_{k=1}^{4j+1}$,
respectively. If this is the case, the requirement ${\rm
rank}\|Q\|=(2j+1)^2$ is equivalent to the condition
\begin{equation}
\label{determinant-product} \Delta_1 \Delta_2 \cdot \ldots \cdot
\Delta_{2j} \ne 0,
\end{equation}
\noindent where $\Delta_q$, $q=1,\dots,2j$ are expressed through
${\bf n}_k =
(\sin\theta_k\cos\varphi_k,\sin\theta_k\sin\varphi_k,\cos\theta_k)$
and associated Legendre polynomials $P_l^{(m)}(x)$ as follows:
\begin{equation}
\Delta_q = \det \left(%
\begin{array}{cccc}
  P_q^{(0)}(\cos\theta_1) & \cdots & P_q^{(q)}(\cos\theta_1)\cos q\varphi_1 & P_q^{(q)}(\cos\theta_1)\sin q\varphi_1 \\
  P_q^{(0)}(\cos\theta_2) & \cdots & P_q^{(q)}(\cos\theta_2)\cos q\varphi_2 & P_q^{(q)}(\cos\theta_2)\sin q\varphi_2 \\
  P_q^{(0)}(\cos\theta_3) & \cdots & P_q^{(q)}(\cos\theta_3)\cos q\varphi_3 & P_q^{(q)}(\cos\theta_3)\sin q\varphi_3 \\
  \cdots & \cdots & \cdots & \cdots \\
  P_q^{(0)}(\cos\theta_{2q}) & \cdots & P_q^{(q)}(\cos\theta_{2q})\cos q\varphi_{2q} & P_q^{(q)}(\cos\theta_{2q})\sin q\varphi_{2q} \\
  P_q^{(0)}(\cos\theta_{2q+1}) & \cdots & P_q^{(q)}(\cos\theta_{2q+1})\cos q\varphi_{2q+1} & P_q^{(q)}(\cos\theta_{2q+1})\sin q\varphi_{2q+1} \\
\end{array}%
\right).
\end{equation}

In the particular case of $q = 1$, we have
\begin{equation}
\Delta_1 = \det\left(%
\begin{array}{ccc}
  P_1^{(0)}(\cos\theta_1) & P_1^{(1)}(\cos\theta_1)\cos\varphi_1 & P_1^{(1)}(\cos\theta_1)\sin\varphi_1 \\
  P_1^{(0)}(\cos\theta_2) & P_1^{(1)}(\cos\theta_2)\cos\varphi_2 & P_1^{(1)}(\cos\theta_2)\sin\varphi_2 \\
  P_1^{(0)}(\cos\theta_3) & P_1^{(1)}(\cos\theta_3)\cos\varphi_3 & P_1^{(1)}(\cos\theta_3)\sin\varphi_3 \\
\end{array}%
\right) = ({\bf n}_1 \cdot [ {\bf n}_2 \times {\bf n}_3 ] ).
\end{equation}

\begin{figure}[t]
\begin{center}
\includegraphics{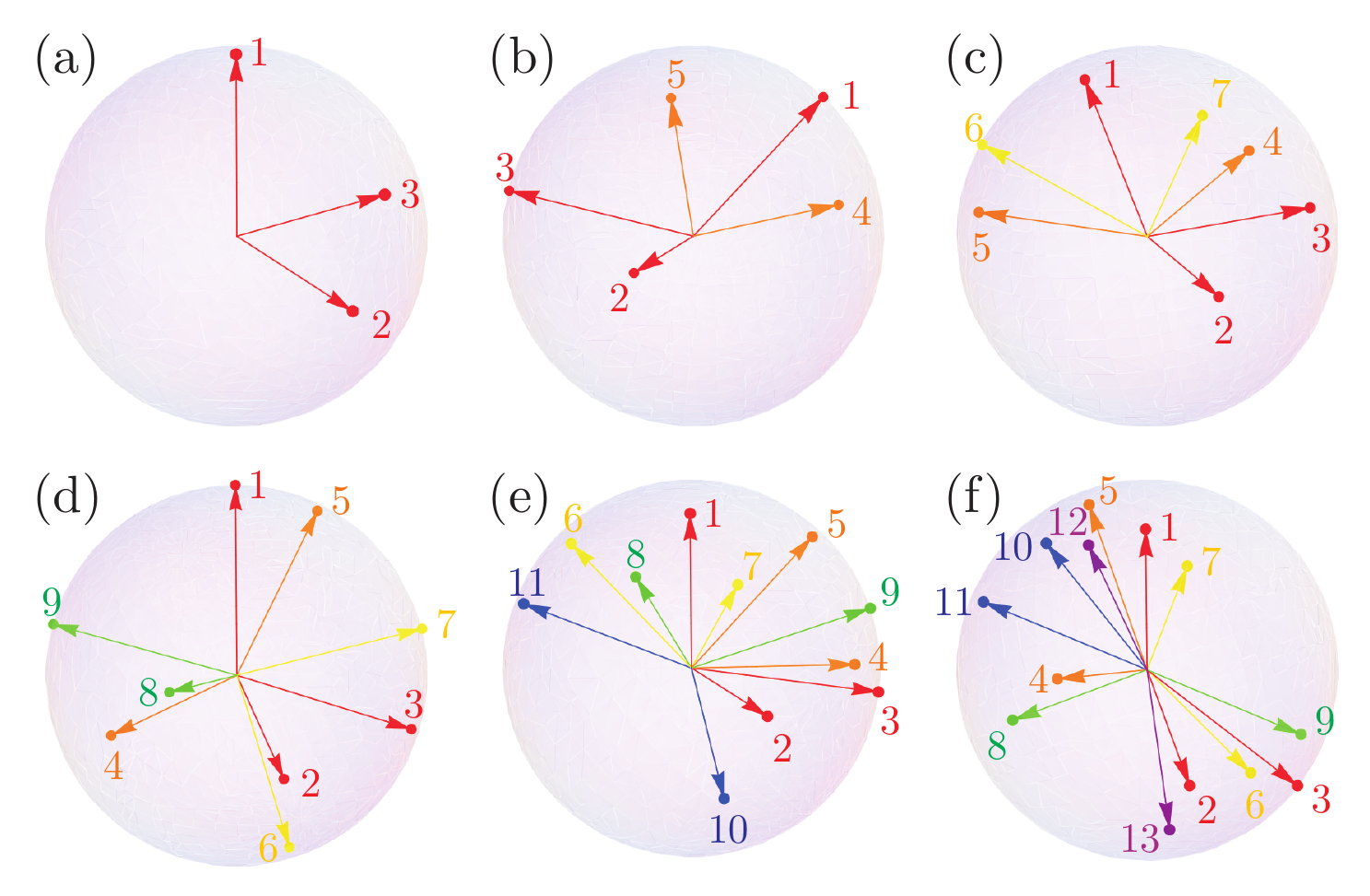}
\caption{\label{figure:optimal-directions} Optimum choice of the
directions $\{{\bf n}_k\}_{k=1}^{4j+1}$ that ensures the minimum
possible errors of the density operator $\hat{\rho}$ with respect
to the errors of the probability vector $\boldsymbol{\cal P}$,
where $j=1/2$~(a), $j=1$~(b), $j=3/2$~(c), $j=2$~(d), $j=5/2$~(e),
and $j=3$~(f).}
\end{center}
\end{figure}

It is worth mentioning that there exists an optimum choice of the
directions $\{{\bf n}_k\}_{k=1}^{4j+1}$ that provides minimum
possible errors of the reconstruction procedure due to errors in
measured probabilities ${\cal P}(m,{\bf n}_k)$. Indeed, according
to the results of computational mathematics (see, e.g.,
\cite{golub}), the errors of the vector $\boldsymbol \rho$ defined
by formula~(\ref{P-map}) are directly proportional to the
condition number $\mu(Q)$ of the matrix $\|Q\|$. The greater the
product $\Delta_1\cdot \ldots \cdot \Delta_{2j}$, the smaller
$\mu(Q)$ and, consequently, the errors of the reconstruction
procedure. For qubits ($j=1/2$), the optimum choice is three
orthogonal vectors $\{{\bf n}_k\}_{k=1}^3$ because their triple
product takes the maximum value in this case. As far as higher
spins are concerned, maximization of
expression~(\ref{determinant-product}) is performed numerically,
and the optimum directions $\{{\bf n}_k\}_{k=1}^{(4j+1)}$ are
shown in Fig.~\ref{figure:optimal-directions}.

%%%%%%%%%%%%%%%%%%%%%%%%%%%%%%%%%%%%%%%%%%%%%%%%%%%%%%%%%%%%%%%%%%%%%%%%%%%%%%%%%
%%%%%%%%%%%%%%%%%%%%%%%%%%%%%%%%%%%%%%%%%%%%%%%%%%%%%%%%%%%%%%%%%%%%%%%%%%%%%%%%%
%%%%%%%%%%%%%%%%%%%%%%%%%%%%%%%%%%%%%%%%%%%%%%%%%%%%%%%%%%%%%%%%%%%%%%%%%%%%%%%%%
%%%%%%%%%%%%%%%%%%%%%%%%%%%%%%%%%%%%%%%%%%%%%%%%%%%%%%%%%%%%%%%%%%%%%%%%%%%%%%%%%

A particular case of
$\theta_1=\theta_2=\ldots=\theta_{4j+1}=\theta$ corresponds to the
Newton--Young reconstruction procedure~\cite{newton} and implies
the limitation $P_L^{(m)}(\cos\theta) \ne 0$ for all $0 \le m \le
L$, $1 \le L \le 2j$. A schematic illustration of this
reconstruction procedure is given in Fig.~\ref{Weigert-Newton}.

An alternative way to meet the requirement ${\rm
rank}\|Q\|=(2j+1)^2$ is to ensure linear independence of vectors
$\{{\bf S}_{L}^{(j)}({\bf n}_k)\}_{k=1}^{2L+1}$ for all
$L=1,\dots,2j$. Linear independence of these vectors is equivalent
to nonzero Gram determinant $\det\left\|\Big({\bf
S}_{L}^{(j)}({\bf n}_i) \cdot {\bf S}_{L}^{(j)}({\bf
n}_k)\Big)\right\|_{i,k=1}^{2L+1}$. Since $\Big({\bf
S}_{L}^{(j)}({\bf n}_i) \cdot {\bf S}_{L}^{(j)}({\bf
n}_k)\Big)=P_L^{(0)}({\bf n}_i \cdot {\bf n}_k)$, we obtain
\begin{equation}
\det\left\|P_1^{(0)}({\bf n}_i \cdot {\bf
n}_k)\right\|_{i,k=1}^{3} \cdot \det\left\|P_2^{(0)}({\bf n}_i
\cdot {\bf n}_k)\right\|_{i,k=1}^{5} \cdot \ldots \cdot
\det\left\|P_{2j}^{(0)}({\bf n}_i \cdot {\bf
n}_k)\right\|_{i,k=1}^{4j+1} \ne 0.
\end{equation}
\noindent In the case of qubits, one has
\begin{equation}
\label{condition-qubits}
\Delta_1 ' = \det\left(%
\begin{array}{ccc}
  ({\bf n}_1\cdot{\bf n}_1) & ({\bf n}_1\cdot{\bf n}_2) & ({\bf n}_1\cdot{\bf n}_3) \\
  ({\bf n}_2\cdot{\bf n}_1) & ({\bf n}_2\cdot{\bf n}_2) & ({\bf n}_2\cdot{\bf n}_3) \\
  ({\bf n}_3\cdot{\bf n}_1) & ({\bf n}_3\cdot{\bf n}_2) & ({\bf n}_3\cdot{\bf n}_3) \\
\end{array}%
\right) = ({\bf n}_1 \cdot [ {\bf n}_2 \times {\bf n}_3 ])^2 \ne
0.
\end{equation}
\noindent As far as qutrits ($j=1$) are concerned, we obtain
\begin{equation}
\label{condition-qutrits}
\Delta_1 ' \! \cdot \! \det \! \left(%
\begin{array}{ccccc}
  1 & \frac{3({\bf n}_1 \cdot {\bf n}_2)^2-1}{2} & \frac{3({\bf n}_1 \cdot {\bf n}_3)^2-1}{2}
  & \frac{3({\bf n}_1 \cdot {\bf n}_4)^2-1}{2} & \frac{3({\bf n}_1 \cdot {\bf n}_5)^2-1}{2} \\
  \frac{3({\bf n}_2 \cdot {\bf n}_1)^2-1}{2} & 1 & \frac{3({\bf n}_2 \cdot {\bf n}_3)^2-1}{2}
  & \frac{3({\bf n}_2 \cdot {\bf n}_4)^2-1}{2} & \frac{3({\bf n}_2 \cdot {\bf n}_5)^2-1}{2} \\
  \frac{3({\bf n}_3 \cdot {\bf n}_1)^2-1}{2} & \frac{3({\bf n}_3 \cdot {\bf n}_2)^2-1}{2} & 1
  & \frac{3({\bf n}_3 \cdot {\bf n}_4)^2-1}{2} & \frac{3({\bf n}_3 \cdot {\bf n}_5)^2-1}{2} \\
  \frac{3({\bf n}_4 \cdot {\bf n}_1)^2-1}{2} & \frac{3({\bf n}_4 \cdot {\bf n}_2)^2-1}{2} &
  \frac{3({\bf n}_4 \cdot {\bf n}_3)^2-1}{2} & 1 & \frac{3({\bf n}_4 \cdot {\bf n}_5)^2-1}{2} \\
  \frac{3({\bf n}_5 \cdot {\bf n}_1)^2-1}{2} & \frac{3({\bf n}_5 \cdot {\bf n}_2)^2-1}{2} &
  \frac{3({\bf n}_5 \cdot {\bf n}_3)^2-1}{2} & \frac{3({\bf n}_5 \cdot {\bf n}_4)^2-1}{2} & 1\\
\end{array}%
\right) \! \ne \! 0.
\end{equation}

\subsection{\label{subsection:number-of-SU(N)} $\textit{SU(N)}$ Rotations in Hilbert Space}

~~~ If the matrix $u$ in the definition of spin tomogram
$w^{(j)}(m,u)$ is an element of the group $SU(N)$ with $2 < N \le
2j+1$, the tomogram $w^{(j)}(m,u)$ is also referred to as a
unitary spin tomogram~\cite{man'ko-sudarshan}. The case $N=2j+1$
corresponds to the most general form of unitary rotations in the
Hilbert space of spin $j$. By the previous statement, we can
assume linear independence of vectors $\{{\bf U}(m, u_k)\}$, where
the spin projection $m$ takes values $m=-j,-j+1,\dots,j-1$ and $k$
runs from $1$ to $N_u$. Here $N_u$ denotes the minimum number of
spin portraits ${\bf w}_j^{(j)}(u_k)$, $u_k\in SU(N)$, $N=2j+1$,
that are needed to construct a bijective map. Since
$\sum_{m=-j}^{j}\hat{U}^{(j)}(m,u_k)=\hat{I}$ for all
$k=1,\dots,N_u$, we have one more independent vector. Thus, the
total number of linear independent vectors $\{{\bf U}(m, u_k)\}$
equals $2j N_u+1$. On the other hand, this number must be equal to
$(2j+1)^2$ for the map to be invertible. Consequently,
$N_u=((2j+1)^2-1)/2j=2j+2$.

We see that one needs fewer $SU(N)$ rotations than $SU(2)$
rotations in order to map a quantum state onto a single
probability distribution. Nevertheless, this advantage is
accompanied by the complexity of the experimental realization of
$SU(N)$ rotations with regard to $SU(2)$ rotations in the Hilbert
space. For intermediate cases $2<N<2j+1$, we have $2j+2 \le N_u <
4j+1$.

Let us consider the case of $SU(N)$ rotations with $N=2j+1$ in
detail. As in the previous section, any quantum state is mapped
onto a single $(2j+1)(2j+2)$-probability vector $\boldsymbol{\cal
P}$ as follows:
\begin{equation}
\label{P-SUN} {\boldsymbol {\cal P}} = \|R\| \boldsymbol\rho,
\end{equation}
\noindent where the $(2j+1)(2j+2)\times (2j+1)^2$ rectangular
matrix $\|R\|$ reads
\begin{equation}
\|R\| = \left(%
\begin{array}{c}
 p_1 {\bf U}^{(j)}(j,u_1)^{T} \\
 p_1 {\bf U}^{(j)}(j-1,u_1)^{T} \\
  \cdots \\
 p_1 {\bf U}^{(j)}(-j,u_1)^{T} \\
 \vdots\\
 p_{2j+2} {\bf U}^{(j)}(j,u_{2j+2})^{T} \\
 p_{2j+2} {\bf U}^{(j)}(j-1,u_{2j+2})^{T} \\
  \cdots \\
 p_{2j+2} {\bf U}^{(j)}(-j,u_{2j+2})^{T} \\
\end{array}%
\right).
\end{equation}
\noindent Taking into account the condition ${\rm rank}
\|R\|=(2j+1)^2$, we express the inverse map $\boldsymbol{\cal P}
\rightarrow \boldsymbol \rho$ through a pseudo-inverse matrix
$\|R^{+}\|$~\cite{gantmacher} as follows:
\begin{equation}
\label{pseudoinverse} \boldsymbol \rho = \|R^{+}\|
\boldsymbol{\cal P} =
\left(\|R^{\dag}\|\|R\|\right)^{-1}\|R^{\dag}\| \boldsymbol{\cal
P}.
\end{equation}
\noindent The requirement ${\rm rank} \|R\|=(2j+1)^2$ is
equivalent to the linear independence of vectors ${\bf
U}^{(j)}(m,u_k)$, $m=-j,-j+1,\dots,j-1$, $k=1,2,\dots,2j+2$ and
the vector ${\bf I}=\sum_{m=-j}^{j}{\bf U}^{(j)}(m,u_k)$, which
does not rely on $k$. The linear independence of these vectors is
achieved whenever the corresponding Gram determinant is nonzero.
That yields the constraint of the form
\begin{equation}
\Gamma = \det \left( \begin{array}{c|c|c|c|c}
  2j+1 & 1 \ 1 \ \dots \ 1 & 1 \ 1 \ \dots \ 1 & \cdots & 1 \ 1 \ \dots \ 1 \\
  \hline
  \begin{smallmatrix}
    1 \\
    1 \\
    \cdots\\
    1
  \end{smallmatrix} & I_{2j} & \Lambda(u_1 u_2^{\dag}) & \cdots & \Lambda(u_1 u_{2j+2}^{\dag}) \\
  \hline
  \begin{smallmatrix}
    1 \\
    1 \\
    \cdots\\
    1
  \end{smallmatrix} & \Lambda(u_2 u_1^{\dag}) & I_{2j} & \cdots & \Lambda(u_2 u_{2j+2}^{\dag}) \\
  \hline
  \cdots & \cdots & \cdots & \cdots & \cdots \\
  \hline
  \begin{smallmatrix}
    1 \\
    1 \\
    \cdots\\
    1
  \end{smallmatrix} & \Lambda(u_{2j+2} u_1^{\dag}) & \Lambda(u_{2j+2} u_2^{\dag}) & \cdots & I_{2j} \\
\end{array} \right) \ne 0,
\end{equation}
\noindent where the Gram matrix is composed of blocks: $~I_{2j}$
is the $2j \times 2j$ unity matrix and $\Lambda(u_k
u_{k'}^{\dag})$ is a $2j \times 2j$ matrix with elements
$\|\Lambda(u_k u_{k'}^{\dag})\|_{ln}=|\langle j l | u_k
u_{k'}^{\dag} | j n \rangle|^2$. Using the orthogonal expansion of
dequantizer~(\ref{quant-orth}), we rewrite the condition obtained
in terms of vectors ${\bf S}_{L}^{(j)}(u_k)$ as follows:
\begin{equation}
\label{Gamma-prime} \Gamma' = \det \left( \begin{array}{c|c|c|c}
  I_{2j} & \Lambda'(u_1,u_2) & \cdots & \Lambda'(u_1,u_{2j+2}) \\
  \hline
  \Lambda'(u_2,u_1) & I_{2j} & \cdots & \Lambda'(u_2,u_{2j+2}) \\
  \hline
  \cdots & \cdots & \cdots & \cdots \\
  \hline
  \Lambda'(u_{2j+2},u_1) & \Lambda'(u_{2j+2},u_2) & \cdots & I_{2j} \\
\end{array} \right) \ne 0,
\end{equation}
\noindent where $\Lambda'(u_k,u_{k'})$ is a $2j \times 2j$ matrix
with elements $\|\Lambda'(u_k,u_{k'})\|_{L L'}=\Big({\bf
S}_{L}^{(j)}(u_k) \cdot {\bf S}_{L'}^{(j)}(u_{k'})\Big)$. Note
that $\Gamma'$ is nothing else but the volume of the
$2j(2j+2)$-dimensional parallelogram with edges ${\bf
S}_{L}^{(j)}(u_k)$, $L=1,\dots,2j$, $k=1,\dots, 2j+2$. Moreover,
$0 \le \Gamma' \le 1$ since all vectors ${\bf S}_{L}^{(j)}(u_k)$
are normalized.

The optimum choice of unitary matrices $u_k \in SU(N)$, $N=2j+1$,
follows the same line of reasoning as in the case of $SU(2)$
rotations. Indeed, if the probability vector $\boldsymbol{\cal P}$
is measured experimentally within the accuracy
$\delta\boldsymbol{\cal P}$, formula~(\ref{pseudoinverse}) yields
the vector $\boldsymbol\rho$ defined with an error bar of
$\delta\boldsymbol\rho$. It is known that
$\frac{\|\delta\boldsymbol\rho\|_2}{\|\boldsymbol\rho\|_2} \le \mu
\frac{\|\delta\boldsymbol{\cal P}\|_2}{\|\boldsymbol{\cal
P}\|_2}$, where $\mu$ is the condition number of the
matrix~(\ref{Gamma-prime}) and $\|\cdot\|_2$ is the Euclidean norm
of a vector. It can be shown that $\mu \le
\frac{1+\sqrt{1-\Gamma'}}{1-\sqrt{1-\Gamma'}}$. Consequently, the
greater $\Gamma'$ (or $\Gamma$), the less erroneous is the
reconstructed state $\boldsymbol\rho$.

\section{\label{section:reconstruction} Inverse Mapping of a Probability Vector onto the Density Matrix}

~~~ In this section, we give an explicit expression of the density
operator $\hat{\rho}$ of the system with spin $j$ in terms of the
single probability vector $\boldsymbol{\cal P}$, which could
itself be treated as the notion of quantum state. We consider
distributions $\boldsymbol{\cal P}$ obtained by $SU(2)$ rotations.
It was shown in Sec.~\ref{section:inverse-qubit-portrait} that any
vector $\boldsymbol{\cal P}$ is readily transformed into the
vector $\boldsymbol{\cal P}_{\rm eq}$. For this reason, we will
focus attention on the map $\boldsymbol{\cal P}_{\rm eq}
\rightarrow \hat{\rho}$.

Let us now recall that the direct map reads
\begin{equation}
\label{P-recall} {\cal P}_{\rm eq}(m,{\bf n}_k) = \frac{1}{4j+1}
w^{(j)}(m,{\bf n}_k) =
  \frac{1}{4j+1} {\rm Tr}\Big( \hat{\rho} ~ \hat{U}^{(j)}(m,{\bf n}_k) \Big),
\end{equation}
\noindent where $\hat{U}^{(j)}(m,{\bf n}_k)$ is the dequantizer
operator that can be resolved into sum~(\ref{dequant-orth}). This
means that
\begin{eqnarray}
\label{P-scanning-expansion} {\cal P}_{\rm eq}(m,{\bf n}_k) =
\frac{1}{4j+1} w^{(j)}(m,{\bf n}_k) =
  \frac{1}{4j+1} \sum_{L=0}^{2j} f_L^{(j)}(m) {\rm Tr}\Big( \hat{\rho}
  \hat{S}_L^{(j)}({\bf n}_k) \Big) = \sum_{L=0}^{2j} {\rm Tr}\Big( \hat{\rho} ~
  \hat{\cal U}_L^{(j)}(m,{\bf n}_k) \Big),\nonumber\\
\end{eqnarray}
\noindent where we introduce the $L$-dequantizer operator
$\hat{\cal U}_L^{(j)}(m,{\bf n}_k) = (4j+1)^{-1} f_L^{(j)}(m)
\hat{S}_L^{(j)}({\bf n}_k)$.

Since the direct map is linear, it can be assumed that the inverse
map is also linear, that is,
\begin{equation}
\label{rho:linear:map} \hat{\rho} =
\sum_{k=1}^{4j+1}\sum_{m=-j}^{j} {\cal P}_{\rm eq}(m,{\bf n}_k)
\hat{\cal D}^{(j)}(m,k),
\end{equation}
\noindent where $\hat{\cal D}^{(j)}(m,k)$ is the quantizer
operator to be determined. We already know that it is convenient
to rearrange directions $\{{\bf n}_k\}_{k=1}^{4j+1}$ and consider
sets $\{{\bf n}_k\}_{k=1}^{2L+1}$, $L=0,1,\dots,2j$. In view of
this fact, we treat a solution of the form
\begin{equation}
\label{rho:L-expansion} \hat{\rho} =
\sum_{L=0}^{2j}\sum_{k=1}^{2L+1}\sum_{m=-j}^{j}
  {\cal P}_{\rm eq}(m,{\bf n}_k) \hat{\cal D}_L^{(j)}(m,k).
\end{equation}
\noindent If $L$-quantizers $\hat{\cal D}_L^{(j)}(m,k)$ are known,
we immediately have $\hat{\cal D}^{(j)}(m,k) = \sum\limits_{L:~
(k-1)/2 \le L\le 2j} \hat{\cal D}_L^{(j)}(m,k)$.

{\bf Proposition}. The $L$-quantizer is expressed through
operators $\hat{S}_L^{(j)}({\bf n}_{k'})$ and Gram matrix
$\|\mathscr{M}(L)\|$, whose matrix elements are
$\|\mathscr{M}(L)\|_{kk'}={\rm Tr}\left(\hat{S}_L^{(j)}({\bf
n}_k)\hat{S}_L^{(j)}({\bf n}_{k'})\right) = \left({\bf
S}_L^{(j)}({\bf n}_k) \cdot {\bf S}_L^{(j)}({\bf n}_{k'})\right) =
P_L({\bf n}_k \cdot {\bf n}_{k'})$, as follows:
\begin{equation}
\label{D-L-quantizer} \hat{\cal D}_L^{(j)}(m,k) = (4j+1)
f_L^{(j)}(m)\sum_{k'=1}^{2L+1}\|\mathscr{M}^{-1}(L)\|_{kk'}
\hat{S}_L^{(j)}({\bf n}_{k'}),
\end{equation}
\noindent with the operators
$\left\{\sum_{k'=1}^{2L+1}\|\mathscr{M}^{-1}(L)\|_{kk'}
\hat{S}_L^{(j)}({\bf n}_{k'})\right\}_{k=1}^{2L+1}$ forming a dual
basis for the given basis\\ $\left\{\hat{S}_L^{(j)}({\bf
n}_k)\right\}_{k=1}^{2L+1}$.

$\square$

Let us check that formula (\ref{D-L-quantizer}) gives an adequate
solution of the problem.

If the directions $\{{\bf n}_k\}_{k=1}^{2j+1}$ are chosen properly
and the requirement (\ref{determinant-product}) is satisfied
(which is equivalent to $\prod_{L=1}^{2j}\det\|\mathscr{M}(L)\|
\ne 0$), then any density operator $\hat{\rho}$ is resolved into a
sum of orthogonal operators
\begin{equation}
\label{rho-in-proof}
\hat{\rho}=\sum_{L'=0}^{2j}\sum_{k'=1}^{2L'+1}a(L',k')\hat{S}_{L'}^{(j)}({\bf
n}_{k'}).
\end{equation}
Substituting formula (\ref{rho-in-proof}) for $\hat{\rho}$ in
(\ref{P-scanning-expansion}), we obtain
\begin{eqnarray}
&& {\rm Tr}\Big( \hat{\rho} \hat{S}_L^{(j)}({\bf n}_k) \Big) =
\sum_{L'=0}^{2j}\sum_{k'=1}^{2L'+1} a(L',k') {\rm Tr}\Big(
\hat{S}_{L'}^{(j)}({\bf n}_{k'}) \hat{S}_L^{(j)}({\bf n}_k) \Big)
= \sum_{k'=1}^{2L+1} a(L,k')
\|\mathscr{M}(L)\|_{kk'}, \\
&& \label{P-m-n_k-inside-proof} {\cal P}_{\rm eq}(m,{\bf n}_k) =
  \frac{1}{4j+1} \sum_{L=0}^{2j} f_L^{(j)}(m) \sum_{k'=1}^{2L+1} a(L,k') \|\mathscr{M}(L)\|_{kk'}.
\end{eqnarray}
After combining (\ref{D-L-quantizer}) and
(\ref{P-m-n_k-inside-proof}), direct calculation of the right-hand
side of Eq.~(\ref{rho:L-expansion}) yields
\begin{eqnarray}
&& \sum_{L=0}^{2j}\sum_{k=1}^{2L+1}\sum_{m=-j}^{j} {\cal P}_{\rm
eq}(m,{\bf
n}_k) \hat{\cal D}_L^{(j)}(m,k) \nonumber\\
&& = \sum_{L,L''=0}^{2j} \sum_{k,k''=1}^{2L+1}
\sum_{k'=1}^{2L''+1} \bigg[
\underbrace{\sum_{m=-j}^{j}f_{L''}^{(j)}(m)f_{L}^{(j)}(m)}_{\delta_{L''L}}
\bigg] a(L'',k') \|\mathscr{M}(L'')\|_{kk'}
\|\mathscr{M}^{-1}(L)\|_{kk''} \hat{S}_L^{(j)}({\bf n}_{k'})
\nonumber \\
&& = \sum_{L=0}^{2j} \sum_{k',k''=1}^{2L+1} a(L,k') \bigg[
\underbrace{\sum_{k=1}^{2L+1} \|\mathscr{M}(L)\|_{kk'}
\|\mathscr{M}^{-1}(L)\|_{kk''}}_{\delta_{k'k''}} \bigg]
\hat{S}_L^{(j)}({\bf n}_{k'}) =
\sum_{L=0}^{2j}\sum_{k'=1}^{2L+1}a(L,k')\hat{S}_{L}^{(j)}({\bf
n}_{k'}) = \hat{\rho}. \nonumber\\
\end{eqnarray}
This concludes the proof.

$\blacksquare$

In fact, the proof above is followed by the relation between the
$L$-dequantizer and the $L'$-quantizer
\begin{equation}
{\rm Tr} \left( \hat{\cal U}_L^{(j)}(m,{\bf n}_k) \hat{\cal
D}_{L'}^{(j)}(m',{\bf n}_{k'}) \right) =
f_L^{(j)}(m)f_L^{(j)}(m')\delta_{LL'}\delta_{kk'}.
\end{equation}

To summarize the results of this section, we write the explicit
form of the quantizer
\begin{equation}
\label{quantizer-k} \hat{\cal D}^{(j)}(m,k) = \sum\limits_{L:~
(k-1)/2 \le L\le 2j} \hat{\cal D}_L^{(j)}(m,k) = (4j+1)
\sum\limits_{L:~ (k-1)/2 \le L\le 2j}
f_L^{(j)}(m)\sum_{k'=1}^{2L+1}\|\mathscr{M}^{-1}(L)\|_{kk'}
\hat{S}_L^{(j)}({\bf n}_{k'}).
\end{equation}

\section{\label{section:kernels} Star-Product and Intertwining Kernels}

~~~ Suppose ${\cal P}_{{\rm eq}~i}(m_i,{\bf n}_{k_i})$ is the
symbol (\ref{P-recall}) of a state $\hat{\rho}_i$, $i=1,2$; then
the operator $\hat{\rho}_1\hat{\rho}_2$ is associated with a
symbol ${\cal P}_{{\rm eq}~3}(m_3,{\bf n}_{k_3})$ which is called
the star
product~\cite{manko:star:1,manko:star:2,omanko-star-brief} of
symbols ${\cal P}_{{\rm eq}~1}(m_1,{\bf n}_{k_1})$ and ${\cal
P}_{{\rm eq}~2}(m_2,{\bf n}_{k_2})$ and denoted by $({\cal
P}_{{\rm eq}~1} \star {\cal P}_{{\rm eq}~2})(m_3,{\bf n}_{k_3})$.
Combining (\ref{P-scanning-expansion}) and (\ref{quantizer-k}), it
is not hard to see that
\begin{eqnarray}
{\cal P}_{{\rm eq}~3}(m_3,{\bf n}_{k_3}) =
\sum_{k_1,k_2=1}^{4j+1}\sum_{m_1,m_2=-j}^{j} K^{(j)}(m_3,{\bf
n}_{k_3},m_2,{\bf n}_{k_2},m_1,{\bf n}_{k_1}) {\cal P}_{{\rm
eq}~2}(m_2,{\bf n}_{k_2}) {\cal P}_{{\rm eq}~1}(m_1,{\bf
n}_{k_1}),\nonumber\\
\end{eqnarray}
\noindent where the star-product kernel $K^{(j)}(m_3,{\bf
n}_{k_3},m_2,{\bf n}_{k_2},m_1,{\bf n}_{k_1})$ reads
\begin{eqnarray}
\label{star-product-kernel} && K^{(j)}(m_3,{\bf n}_{k_3},m_2,{\bf
n}_{k_2},m_1,{\bf n}_{k_1}) =
  {\rm Tr} \left[ \hat{\cal D}^{(j)}(m_1,{n}_{k_1}) \hat{\cal D}^{(j)}(m_2,{n}_{k_2})
\hat{\cal U}^{(j)}(m_3,{n}_{k_3}) \right] \nonumber\\
&& = (4j+1) \sum\limits_{\{L_1:~ (k_1-1)/2 \le L_1 \le 2j\}}
  \sum\limits_{\{L_2:~ (k_2-1)/2 \le L_2 \le 2j\}}
  \sum_{L_3=0}^{2j} f_{L_1}^{(j)}(m_1) f_{L_2}^{(j)}(m_2) f_{L_3}^{(j)}(m_3)
\nonumber\\
&& \quad \times \sum_{k_1'=1}^{2L_1+1} \sum_{k_2'=1}^{2L_2+1}
  \|\mathscr{M}^{-1}(L_1)\|_{k_1k_1'} \|\mathscr{M}^{-1}(L_2)\|_{k_2k_2'}
  {\rm Tr} \left[ \hat{S}_{L_1}^{(j)}({\bf n}_{k_1'}) \hat{S}_{L_2}^{(j)}({\bf n}_{k_2'})
  \hat{S}_{L_3}^{(j)}({\bf n}_{k_3})\right].
\end{eqnarray}

Let us recall that we have considered previously two maps of the
density operator $\hat{\rho}$ onto the probability distribution
functions, namely, the map onto tomograms $w^{(j)}(m,{\bf n})$
depending on the continuous variable ${\bf n}={\bf
n}(\theta,\varphi)\in S^2$ and the map onto the single probability
distribution ${\cal P}_{\rm eq}(m,{\bf n}_k)$ depending on the
discrete variable ${\bf n}_k$, $k=1,2,\dots, 4j+1$. Since both
maps are invertible, symbols $w^{(j)}(m,{\bf n})$ and ${\cal
P}_{\rm eq}(m,{\bf n}_k)$ are related by intertwining kernels.
Indeed, combining formulas~(\ref{rho-reconstr-tomographic}) and
(\ref{P-recall}), we obtain
\begin{equation}
{\cal P}_{\rm eq}(m,{\bf n}_k) = \sum_{m'=-j}^{j}
\int\limits_{S^2} \frac{d{\bf n}'}{4\pi} K_{w\rightarrow{\cal
P}}^{(j)}(m,{\bf n}_k,m',{\bf n}') w^{(j)}(m',{\bf n}'),
\end{equation}
\noindent where $K_{w\rightarrow{\cal P}}^{(j)}(m,{\bf
n}_k,m',{\bf n}') = (4j+1)^{-1} {\rm Tr} \left(
\hat{D}^{(j)}(m',{\bf n}')\hat{U}^{(j)}(m,{\bf n}_k) \right)$.
Using expansions~(\ref{dequant-orth}) and (\ref{quant-orth}) along
with the orthogonality property ${\rm Tr}\left( \hat{S}_{L'}({\bf
n}') \hat{S}_{L}({\bf n}_k) \right)= \delta_{LL'} P_L({\bf
n}'\cdot{\bf n}_k)$, we arrive at
\begin{equation}
\label{intertwining-w-P} K_{w\rightarrow{\cal P}}^{(j)}(m,{\bf
n}_k,m',{\bf n}') = (4j+1)^{-1} \sum_{L=0}^{2j} (2L+1)
f_L^{(j)}(m') f_L^{(j)}(m) P_L({\bf n}'\cdot{\bf n}_k).
\end{equation}

In view of the same argument, the tomogram $w^{(j)}(m,{\bf n})$ is
expressed through the joint probability distribution ${\cal
P}_{\rm eq}(m',{\bf n}_{k'})$ as follows:
\begin{equation}
w^{(j)}(m,{\bf n}) = \sum_{k'=1}^{4j+1}\sum_{m'=-j}^{j} K_{{\cal
P}\rightarrow w}^{(j)}(m,{\bf n},m',k')
  {\cal P}_{\rm eq}(m',{\bf n}_{k'}),
\end{equation}
\noindent where $K_{{\cal P}\rightarrow w}^{(j)}(m,{\bf n},m',k')
= {\rm Tr} \left( \hat{\cal D}^{(j)}(m',k')\hat{U}^{(j)}(m,{\bf
n}) \right)$. Taking into account explicit
formula~(\ref{quantizer-k}), we obtain
\begin{equation}
\label{intertwining-P-w} K_{{\cal P}\rightarrow w}^{(j)}(m,{\bf
n},m',k') = (4j+1) \sum\limits_{L:~ (k'-1)/2 \le L\le 2j}
f_L^{(j)}(m') f_L^{(j)}(m)
\sum_{k=1}^{2L+1}\|\mathscr{M}^{-1}(L)\|_{k'k} P_L({\bf
n}_{k}\cdot{\bf n}).
\end{equation}

\section{\label{section:examples} Examples: Qubits and Qutrits}

~~~ In this section, the results of the previous sections are
specified for two particular cases of lowest spins, namely, qubits
($j=1/2$) and qutrits ($j=1$). As far as qubits are concerned,
only the $SU(2)$ rotations are possible. A quantum state is
associated with the six-dimensional probability vector
$\boldsymbol{\cal P}$ with components ${\cal P}(m,{\bf n}_k)$, $m=
\pm 1/2$ and $k=1,2,3$. In other words, the probability vector
$\boldsymbol{\cal P}$ is composed of three qubit portraits defined
by directions $\{{\bf n}_k\}_{k=1}^{3}$. If these directions are
equiprobable (chosen with the same probability $p_k=1/3$), the
corresponding probability vector is denoted as $\boldsymbol{\cal
P}_{\rm eq}$. Note that formula~(\ref{tom-from-P}) defines the
mapping $\boldsymbol{\cal P}\rightarrow \boldsymbol{\cal P}_{\rm
eq}$ for any vector $\boldsymbol{\cal P}$. For the map
$\hat{\rho}\rightarrow \boldsymbol{\cal P}$ to be invertible, the
limitation $({\bf n}_1 \cdot [{\bf n}_2 \times {\bf n}_3]) \ne 0$
is imposed. The least erroneous reconstruction procedure (see
Fig.~\ref{figure:optimal-directions}) takes place if all three
directions are orthogonal, i.e., $({\bf n}_1 \cdot [{\bf n}_2
\times {\bf n}_3]) = \pm 1$. In general, the inverse
map~(\ref{rho:L-expansion}) of the probability vector
$\boldsymbol{\cal P}_{\rm eq}$ onto the density operator
$\hat{\rho}$ reads

\begin{equation}
\label{qubit-reconstruction} \hat{\rho} = \frac{1}{2} \left[ {\cal
P}_{\rm eq}(+1/2,{\bf n}_1) +
  {\cal P}_{\rm eq}(-1/2,{\bf n}_1) \right] \hat{I} +
  \sum_{k=1}^{3}\left[ {\cal P}_{\rm eq}(+1/2,{\bf n}_k) -
  {\cal P}_{\rm eq}(-1/2,{\bf n}_k) \right] (\hat{\boldsymbol\sigma} \cdot {\bf l}_k),
\end{equation}
\noindent where
$\hat{\boldsymbol\sigma}=(\hat{\sigma}_x,\hat{\sigma}_y,\hat{\sigma}_z)$
are the Pauli operators, and the vectors ${\bf l}_k =
\sum_{k'=1}^{3}\|\mathscr{M}^{-1}(L=1)\|_{kk'}{\bf n}_{k'}$.
Direct calculation provides
\begin{equation}
{\bf l}_1 = \frac{[{\bf n}_2 \cdot {\bf n}_3]}{({\bf n}_1 \cdot
[{\bf n}_2 \times {\bf n}_3])}, \quad {\bf l}_2 = \frac{[{\bf n}_3
\cdot {\bf n}_1]}{({\bf n}_1 \cdot [{\bf n}_2 \times {\bf n}_3])},
\quad {\bf l}_3 = \frac{[{\bf n}_1 \cdot {\bf n}_2]}{({\bf n}_1
\cdot [{\bf n}_2 \times {\bf n}_3])}.
\end{equation}

%%%%%%%%%%%%%%%%%%%%%%%%%%%%%%%%%%%%%%%%%%%%%%%%%%%%%%%%%%%%%%%%%%%%%%%%%%%%%%%%%
%%%%%%%%%%%%%%%%%%%%%%%%%%%%%%%%%%%%%%%%%%%%%%%%%%%%%%%%%%%%%%%%%%%%%%%%%%%%%%%%%
\begin{figure}[t]
\begin{center}
\includegraphics{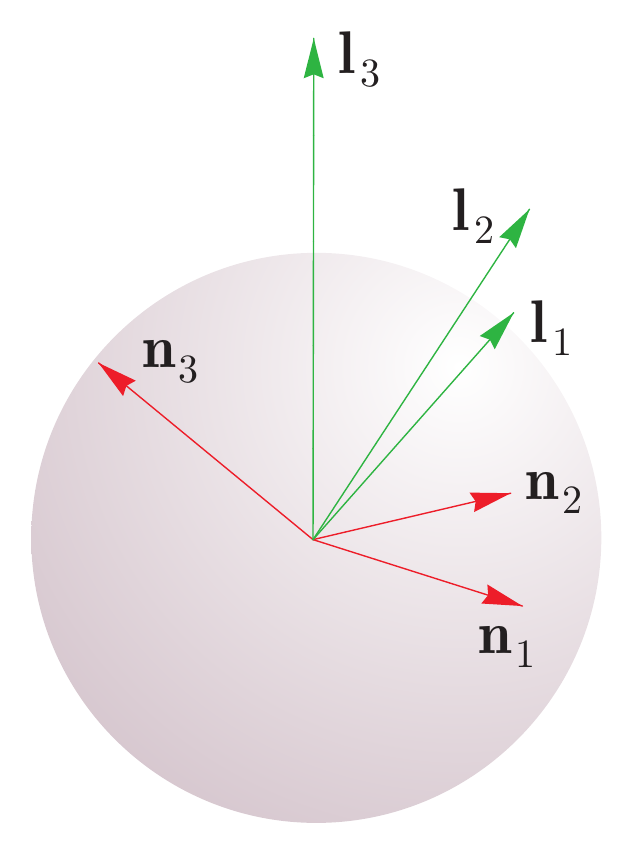}
\caption{\label{figure:vector-product} Duality of basic sets
$\{{\bf n}_k\}_{k=1}^{3}$ and $\{{\bf l}_k\}_{k=1}^{3}$. Any qubit
state is identified with a six-dimensional probability vector
$\boldsymbol{\cal P}$, which is composed of three spin-$1/2$
portraits ${\bf w}_{1/2}^{(1/2)}({\bf n}_1)$, ${\bf
w}_{1/2}^{(1/2)}({\bf n}_2)$, and ${\bf w}_{1/2}^{(1/2)}({\bf
n}_3)$ given by vectors ${\bf n}_1$, ${\bf n}_2$, and ${\bf n}_3$,
respectively. Vectors $\{{\bf l}_k\}_{k=1}^{3}$ form a dual vector
basis that determines the inverse map~(\ref{qubit-reconstruction})
of the vector $\boldsymbol{\cal P}$ onto a qubit density operator.
The map is bijective whenever vectors ${\bf n}_1$, ${\bf n}_2$,
and ${\bf n}_3$ are noncoplanar.}
\end{center}
\end{figure}
%%%%%%%%%%%%%%%%%%%%%%%%%%%%%%%%%%%%%%%%%%%%%%%%%%%%%%%%%%%%%%%%%%%%%%%%%%%%%%%%%
%%%%%%%%%%%%%%%%%%%%%%%%%%%%%%%%%%%%%%%%%%%%%%%%%%%%%%%%%%%%%%%%%%%%%%%%%%%%%%%%%

\noindent Thus, the vectors $\{{\bf l}_k\}_{k=1}^{3}$ form a dual
basis with respect to the directions $\{{\bf
n}_{k'}\}_{k'=1}^{3}$, i.e., $({\bf l}_k \cdot {\bf
n}_{k'})=\delta_{kk'}$. This dual basis can be used to construct
dual symbols of operators~\cite{vitale::dual,omanko:vitale}. Note
that vectors $\{{\bf l}_k\}_{k=1}^{3}$ are no longer normalized.
Figure~\ref{figure:vector-product} illustrates the duality of
basic sets $\{{\bf n}_k\}_{k=1}^{3}$ and $\{{\bf l}_k\}_{k=1}^{3}$
and, consequently, the duality of mappings $\hat{\rho}\rightarrow
\boldsymbol{\cal P}$ and $\boldsymbol{\cal P} \rightarrow
\hat{\rho}$.

Using the special properties of Pauli matrices, one can easily
calculate the star-product kernel~(\ref{star-product-kernel}). The
result is
\begin{eqnarray}
K^{(1/2)}(m_3,{\bf n}_{k_3},m_2,{\bf n}_{k_2},m_1,{\bf n}_{k_1})
&=& 3 \Big\{ \frac{1}{4}\delta_{k_3,1}\delta_{k_2,1} +
\delta_{k_3,1} m_2 m_1 ({\bf l}_{k_2} \cdot {\bf n}_{k_1}) +
\delta_{k_2,1} m_3 m_1 ({\bf l}_{k_3} \cdot {\bf n}_{k_1})
\nonumber\\
&& \quad + m_3 m_2 ({\bf l}_{k_3} \cdot {\bf l}_{k_2}) + 2i m_3
m_2 m_1 ([{\bf l}_{k_3} \times {\bf l}_{k_2}] \cdot {\bf n}_{k_1})
\Big\}.
\end{eqnarray}
\noindent The intertwining kernels (\ref{intertwining-w-P}) and
(\ref{intertwining-P-w}) take the following form:
\begin{eqnarray}
K_{w\rightarrow{\cal P}}^{(1/2)}(m,{\bf n}_k,m',{\bf n}') =
\frac{1}{6} + 2m'm ({\bf n}' \cdot {\bf n}_k), \qquad K_{{\cal
P}\rightarrow w}^{(1/2)}(m,{\bf n},m',k') = 3 \left\{
\frac{1}{2}\delta_{k',1} + 2m'm({\bf l}_{k'}\cdot{\bf n})
\right\}.\nonumber\\
\end{eqnarray}

As far as qutrits are concerned, any quantum state can be
associated either with the fifteen-dimensional probability
vector~(\ref{P-map}) parameterized by five $SU(2)$ rotations or
the twelve-dimensional probability vector (\ref{P-SUN}) written in
terms of four $SU(3)$ rotations in the Hilbert space. The former
case implies that the vector $\boldsymbol{\cal P}$ comprises five
qutrit portraits, each defined by the direction ${\bf n}_k$,
$k=1,\dots,5$. The density operator is uniquely determined
whenever these directions satisfy the
condition~(\ref{condition-qutrits}). The latter case of $SU(3)$
rotations implies the limitation~(\ref{Gamma-prime}) on unitary
matrices $u_{k'}$, $k'=1,\dots,4$.

The explicit formula of the density operator in terms of
experimentally attainable probabilities ${\cal P}_{\rm eq}(m,{\bf
n}_k)$ reads
\begin{eqnarray}
&& \hat{\rho} = \frac{1}{3} \left[ {\cal P}_{\rm eq}(+1,{\bf n}_1)
+
  {\cal P}_{\rm eq}(0,{\bf n}_1) + {\cal P}_{\rm eq}(-1,{\bf n}_1) \right] \hat{I} +
  \frac{1}{2}\sum_{k=1}^{3} \left[ {\cal P}_{\rm eq}(+1,{\bf n}_k) -
  {\cal P}_{\rm eq}(-1,{\bf n}_k) \right] (\hat{\bf J} \cdot
{\bf l}_k)\nonumber \\
&& + \frac{1}{6} \left(%
\begin{array}{c}
  {\cal P}_{\rm eq}(+1,{\bf n}_1) - 2{\cal P}_{\rm eq}(0,{\bf n}_1) + {\cal P}_{\rm eq}(-1,{\bf n}_1) \\
  {\cal P}_{\rm eq}(+1,{\bf n}_2) - 2{\cal P}_{\rm eq}(0,{\bf n}_2) + {\cal P}_{\rm eq}(-1,{\bf n}_2) \\
  {\cal P}_{\rm eq}(+1,{\bf n}_3) - 2{\cal P}_{\rm eq}(0,{\bf n}_3) + {\cal P}_{\rm eq}(-1,{\bf n}_3) \\
  {\cal P}_{\rm eq}(+1,{\bf n}_4) - 2{\cal P}_{\rm eq}(0,{\bf n}_4) + {\cal P}_{\rm eq}(-1,{\bf n}_4) \\
  {\cal P}_{\rm eq}(+1,{\bf n}_5) - 2{\cal P}_{\rm eq}(0,{\bf n}_5) + {\cal P}_{\rm eq}(-1,{\bf n}_5) \\
\end{array}%
\right)^{T} \nonumber\\
&& \times \left\|%
\begin{array}{ccccc}
  1 & \frac{3({\bf n}_1 \cdot {\bf n}_2)^2-1}{2} & \frac{3({\bf n}_1 \cdot {\bf n}_3)^2-1}{2} &
  \frac{3({\bf n}_1 \cdot {\bf n}_4)^2-1}{2} & \frac{3({\bf n}_1 \cdot {\bf n}_5)^2-1}{2} \\
  \frac{3({\bf n}_2 \cdot {\bf n}_1)^2-1}{2} & 1 & \frac{3({\bf n}_2 \cdot {\bf n}_3)^2-1}{2} &
  \frac{3({\bf n}_2 \cdot {\bf n}_4)^2-1}{2} & \frac{3({\bf n}_2 \cdot {\bf n}_5)^2-1}{2} \\
  \frac{3({\bf n}_3 \cdot {\bf n}_1)^2-1}{2} & \frac{3({\bf n}_3 \cdot {\bf n}_2)^2-1}{2} & 1 &
  \frac{3({\bf n}_3 \cdot {\bf n}_4)^2-1}{2} & \frac{3({\bf n}_3 \cdot {\bf n}_5)^2-1}{2} \\
  \frac{3({\bf n}_4 \cdot {\bf n}_1)^2-1}{2} & \frac{3({\bf n}_4 \cdot {\bf n}_2)^2-1}{2} &
  \frac{3({\bf n}_4 \cdot {\bf n}_3)^2-1}{2} & 1 & \frac{3({\bf n}_4 \cdot {\bf n}_5)^2-1}{2} \\
  \frac{3({\bf n}_5 \cdot {\bf n}_1)^2-1}{2} & \frac{3({\bf n}_5 \cdot {\bf n}_2)^2-1}{2} &
  \frac{3({\bf n}_5 \cdot {\bf n}_3)^2-1}{2} & \frac{3({\bf n}_5 \cdot {\bf n}_4)^2-1}{2} & 1\\
\end{array}%
\right\|^{-1} \left(%
\begin{array}{c}
  3(\hat{\bf J}\cdot {\bf n}_1)^2 -2\hat{I} \\
  3(\hat{\bf J}\cdot {\bf n}_2)^2 -2\hat{I} \\
  3(\hat{\bf J}\cdot {\bf n}_3)^2 -2\hat{I} \\
  3(\hat{\bf J}\cdot {\bf n}_4)^2 -2\hat{I} \\
  3(\hat{\bf J}\cdot {\bf n}_5)^2 -2\hat{I} \\
\end{array}%
\right).\nonumber\\
&&
\end{eqnarray}

A similar reconstruction of qutrit states is proposed in
\cite{hofmann}. Higher spins ($S=4$) are reconstructed in
\cite{klose}. The advantage of the proposed inverse
mapping~(\ref{rho:L-expansion}) is that it can be applied to a
system with an arbitrary spin $j$ and provides the explicit
solution in an operator form.

\section{\label{section:quantum-states} Quantum States on $\textit{2j(4j+3)}$-Simplex}

~~~ We already know that any quantum state can be associated with
the probability-distribution vector $\boldsymbol{\cal P}$. If
$SU(2)$ rotations underlie the construction of the vector
$\boldsymbol{\cal P}$, this vector comprises $(2j+1)(4j+1)$
components ${\cal P}(m,{\bf n}_k)$. Consequently, it is
represented by a point on the simplex with dimension
$(2j+1)(4j+1)-1=2j(4j+3)$. Conversely, not all points on the
$2j(4j+3)$ simplex can be associated with quantum states. Indeed,
the condition $\hat{\rho} \ge 0$ is to be satisfied. Let us
reformulate this requirement in terms of components ${\cal
P}(m,{\bf n}_k)$.

In view of the inverse map (\ref{rho:L-expansion}), we readily
obtain the following condition:
\begin{equation}
\label{positivity-requirement}
\sum_{L=0}^{2j}\sum_{k=1}^{2L+1}\sum_{m=-j}^{j}
  {\cal P}_{\rm eq}(m,{\bf n}_k) f_L^{(j)}(m)\sum_{k'=1}^{2L+1}\|\mathscr{M}^{-1}(L)\|_{kk'}
  \hat{S}_L^{(j)}({\bf n}_{k'}) \ge 0.
\end{equation}

%%%%%%%%%%%%%%%%%%%%%%%%%%%%%%%%%%%%%%%%%%%%%%%%%%%%%%%%%%%%%%%%%%%%%%%%%%%%%%%%%
%%%%%%%%%%%%%%%%%%%%%%%%%%%%%%%%%%%%%%%%%%%%%%%%%%%%%%%%%%%%%%%%%%%%%%%%%%%%%%%%%
\begin{figure}[t]
\begin{center}
\includegraphics{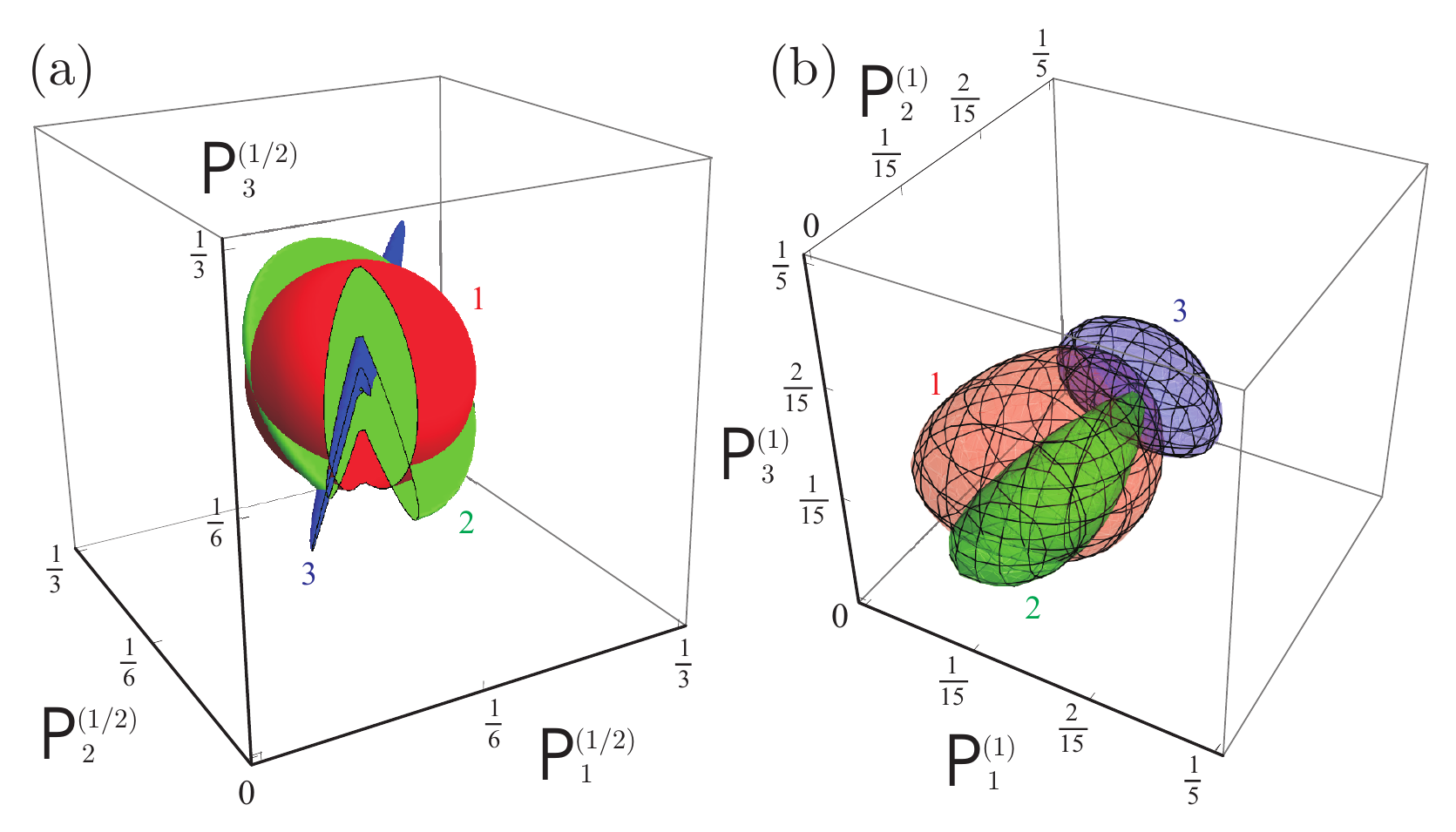}
\caption{\label{figure:qubit-qutrit-states} Convex subset of
quantum states on the $2j(4j+3)$-simplex. In the case of qubits,
${\cal P}_{\rm eq}(+1/2,{\bf n}_1) = \mathsf{P}_{1}^{(1/2)}$,
${\cal P}_{\rm eq}(-1/2,{\bf n}_1) = 1/3-\mathsf{P}_{1}^{(1/2)}$,
${\cal P}_{\rm eq}(+1/2,{\bf n}_2) = \mathsf{P}_{2}^{(1/2)}$,
${\cal P}_{\rm eq}(-1/2,{\bf n}_2) = 1/3-\mathsf{P}_{2}^{(1/2)}$,
${\cal P}_{\rm eq}(+1/2,{\bf n}_3) = \mathsf{P}_{3}^{(1/2)}$, and
${\cal P}_{\rm eq}(-1/2,{\bf n}_3) = 1/3-\mathsf{P}_{3}^{(1/2)}$
(a). So any point on the $\boldsymbol{\cal P}$-simplex is uniquely
determined by a point $\left( \mathsf{P}_{1}^{(1/2)},
\mathsf{P}_{2}^{(1/2)}, \mathsf{P}_{3}^{(1/2)} \right)$ inside a
cube $\{0 \le \mathsf{P}_{k}^{(1/2)} \le 1/3\}_{k=1}^{3}$, where
quantum subsets are shown for $({\bf n}_1\cdot[{\bf n}_2\times
{\bf n}_3])=1$~(1), $0.44$~(2), and $0.02$~(3). In the case of
qutrits, simplex is fourteen-dimensional~(b), where we fix five
directions in such a way that $({\bf n}_1 \cdot {\bf n}_2) = ({\bf
n}_2 \cdot {\bf n}_3) = ({\bf n}_3 \cdot {\bf n}_1) = ({\bf n}_1
\cdot {\bf n}_4) = ({\bf n}_2 \cdot {\bf n}_4) = ({\bf n}_1 \cdot
{\bf n}_5) = ({\bf n}_3 \cdot {\bf n}_5) = 1/\sqrt{3}$. Denote
$\mathsf{P}_{k}^{(1)}={\cal P}_{\rm eq}(+1,{\bf n}_k)$, $k=1,2,3$,
then the cut set ${\cal P}_{\rm eq}(+1,{\bf n}_4)={\cal P}_{\rm
eq}(+1,{\bf n}_5)=1/15$, and ${\cal P}_{\rm eq}(-1,{\bf
n}_{k'})=1/15$, ${k'}=1,\dots,5$ corresponds to 1, the cut set
${\cal P}_{\rm eq}(+1,{\bf n}_4)={\cal P}_{\rm eq}(+1,{\bf
n}_5)=1/15$, and ${\cal P}_{\rm eq}(-1,{\bf n}_{k'})=1/15$,
${k'}=1,\dots,5$ corresponds to 2, and the cut set ${\cal P}_{\rm
eq}(+1,{\bf n}_4)={\cal P}_{\rm eq}(+1,{\bf n}_5)=1/20$, and
${\cal P}_{\rm eq}(-1,{\bf n}_{k'})=1/20$, ${k'}=1,\dots,5$
corresponds to 3.}
\end{center}
\end{figure}
%%%%%%%%%%%%%%%%%%%%%%%%%%%%%%%%%%%%%%%%%%%%%%%%%%%%%%%%%%%%%%%%%%%%%%%%%%%%%%%%%
%%%%%%%%%%%%%%%%%%%%%%%%%%%%%%%%%%%%%%%%%%%%%%%%%%%%%%%%%%%%%%%%%%%%%%%%%%%%%%%%%

\noindent This implies that the matrix of the
operator~(\ref{positivity-requirement}) in the basis of states
$|jm\rangle$ is nonnegative. Nonnegativity of such a matrix is
easily checked by Sylvester's criterion~\cite{meyer}; namely, it
is necessary and sufficient that all principal minors of
matrix~(\ref{positivity-requirement}) are nonnegative.

Let us consider the case of qubits ($j=1/2$) in detail.

From explicit formula (\ref{qubit-reconstruction}), it is readily
seen that vector $\boldsymbol{\cal P}_{\rm eq}$ determines a
quantum state iff
\begin{equation}
\label{simplex-qubit} \left\{ \begin{array}{c}
  \frac{1}{2}[{\cal P}_{\rm eq}(+1/2,{\bf n}_1)+{\cal P}_{\rm eq}(-1/2,{\bf n}_1)] +
    \sum_{k=1}^{3}[{\cal P}_{\rm eq}(+1/2,{\bf n}_k)-{\cal P}_{\rm eq}(-1/2,{\bf n}_k)]
    \big({\bf l}_k \cdot (0,0,1)\big) \ge 0, \\
  \frac{1}{4}[{\cal P}_{\rm eq}(+1/2,{\bf n}_1)+{\cal P}_{\rm eq}(-1/2,{\bf n}_1)]^2 -
    \left(\sum_{k=1}^{3}[{\cal P}_{\rm eq}(+1/2,{\bf n}_k)-{\cal P}_{\rm eq}(-1/2,{\bf n}_k)]
    {\bf l}_k \right)^2 \ge 0, \\
  \frac{1}{2}[{\cal P}_{\rm eq}(+1/2,{\bf n}_1)+{\cal P}_{\rm eq}(-1/2,{\bf n}_1)] -
    \sum_{k=1}^{3}[{\cal P}_{\rm eq}(+1/2,{\bf n}_k)-{\cal P}_{\rm eq}(-1/2,{\bf n}_k)]
    \big({\bf l}_k \cdot (0,0,1)\big) \ge 0. \\
\end{array} \right.
\end{equation}
Taking into account the relation ${\cal P}_{\rm eq}(+1/2,{\bf
n}_k)+
  {\cal P}_{\rm eq}(-1/2,{\bf n}_k)=1/3$, $k=1,2,3$, the obtained system of
inequalities can be easily depicted
(Fig.~\ref{figure:qubit-qutrit-states}a). Indeed, using such a
constrain on the probabilities, the five-simplex for six-component
vector $\boldsymbol{\cal P}$ is identified with the interior of
the cube $0 \le {\cal P}_{\rm eq}(+1/2,{\bf n}_k) \le 1/3$,
$k=1,2,3$. Quantum states are those points on the simplex that
satisfy the conditions~(\ref{simplex-qubit}). In particular, if
$\{{\bf n}_k\}_{k=1}^{3}$ form an orthonormal basis in
$\mathbb{R}^3$, the quantum states are associated with the ball
\begin{equation}
\left( {\cal P}_{\rm eq}(+1/2,{\bf n}_1) - \frac{1}{6} \right)^2 +
  \left( {\cal P}_{\rm eq}(+1/2,{\bf n}_2) - \frac{1}{6} \right)^2 +
  \left( {\cal P}_{\rm eq}(+1/2,{\bf n}_3) - \frac{1}{6} \right)^2 \le \frac{1}{144}.
\end{equation}
\noindent The general case of arbitrary directions $\{{\bf
n}_k\}_{k=1}^{3}$ is shown in Fig.~4a.\footnote{Dramatic
visualization of qubit states in the probability space is
available at Wolfram Demonstrations Project:
\href{http://demonstrations.wolfram.com/RepresentationOfQubitStatesByProbabilityVectors/}{http://demonstrations.wolfram.com/RepresentationOfQubitStatesByProbabilityVectors/}}

In the case of qutrits ($j=1$), we restrict ourselves to a
numerical solution of the system of seven inequalities analogous
to (\ref{simplex-qubit}). It is worth noting that the quantum
domain on the fourteen-simplex is given by algebraic inequalities
--- three inequalities of the first order, three inequalities of
the second order, and one inequality of the third order. Different
cut sets of this simplex by hyperplanes ${\cal P}_{\rm eq}(-1,{\bf
n}_4)= {\cal P}_{\rm eq}(-1,{\bf n}_5)={\rm const}$ and ${\cal
P}_{\rm eq}(0,{\bf n}_k)={\rm const}$, $k=1,\dots,5$ are
illustrated in Fig.~4b.\footnote{The domain of qutrit states in
the probability simplex as well as the dynamics of the condition
number $\mu$ with regard to an arbitrary choice of directions
$\{{\bf n}_k\}_{k=1}^{5}$ is visualized at Wolfram Demonstrations
Project:
\href{http://demonstrations.wolfram.com/QutritStatesAsProbabilityVectors/}{http://demonstrations.wolfram.com/QutritStatesAsProbabilityVectors/}}
The cut set of qutrit states is a third-degree body of the vector
elements ${\cal P}_{\rm eq}(+1,{\bf n}_1)$, ${\cal P}_{\rm
eq}(+1,{\bf n}_2)$, and ${\cal P}_{\rm eq}(+1,{\bf n}_3)$, with
the cut set being located between three planes and three
second-degree surfaces.

\section{\label{section:conclusions}Conclusions}

~~~ To conclude, a bijective map of qudit-$j$ states onto single
probability vectors $\boldsymbol{\cal P}$ has been developed. In
fact, any quantum state is associated with such a probability
vector. Quantum states form a convex subset on a simplex of
possible probability vectors $\boldsymbol{\cal P}$, with the
boundary of quantum states being the $(2j+1)$-degree body of
vector elements ${\cal P}(m,u)$. Examples of quantum subsets are
presented for qubits ($j=1/2$) and qutrits ($j=1$). Components
${\cal P}(m,u)$ are fair probabilities, have a clear physical
meaning, and can be relatively easily measured experimentally.

To be precise, ${\cal P}(m,u_k)$ is a joint probability
distribution function of two discrete variables -- spin projection
$m$ and unitary rotation $u_k$ from a finite set of rotations
$\{u_k\}_{k=1}^{N_u}$. The number of rotations $N_u$ is shown to
depend on the type of rotations used. Namely, $N_u=2j+2$ if all
unitary matrices $u_k$ are elements of the group $SU(N)$ with
$N=2j+1$, and $N_u=4j+1$ if $u_k\in SU(2)$ for all $k$. The latter
case is considered in detail. The dequantizer operator $\hat{\cal
U}^{(j)}(m,{\bf n}_k)$ specifying the direct map
$\hat{\rho}\rightarrow \boldsymbol{\cal P}_{\rm eq}$ and the
quantizer operator $\hat{\cal D}^{(j)}(m,{\bf n}_k)$ specifying
the inverse map $\boldsymbol{\cal P}_{\rm eq} \rightarrow
\hat{\rho}$ are presented in the explicit form for an arbitrary
choice of directions $\{{\bf n}_k\}_{k=1}^{4j+1}$. The kernel of
the corresponding star-product quantization scheme as well as
intertwining kernels relating ${\cal P}$-representation and
$w$-tomographic representation are found.

A subsidiary problem of the optimum choice of directions $\{{\bf
n}_k\}_{k=1}^{4j+1}$ is discussed and partially solved for the low
spin states, with the optimality implying the minimum relative
error $\|\delta \hat{\rho}\|_2$ if errors $\delta \boldsymbol{\cal
P}$ in the measured probability vector $\boldsymbol{\cal P}$ are
presented.

Last, but not least, different mappings of density operators onto
the probability vectors are unified within the concept of the
inverse spin-$s$ portrait method. The difference between mappings
reduces to a particular choice of spin-$s$ portraits implicitly
used in these transforms; namely,
\cite{amiet-weigert-JPA,fuchs-oct-2009} rely on spin-$1/2$
portraits, whereas \cite{newton,hofmann} extensively employ
spin-$j$ portraits.

\section*{Acknowledgments}
~~~ This study was supported by the Russian Foundation for Basic
Research under Project No.~09-02-00142. S.N.F. thanks the Ministry
of Education and Science of the Russian Federation and the Federal
Education Agency for the support under Project No.~2.1.1/5909.

\end{document}